\documentclass{jfm}
\usepackage{amsmath}
\usepackage{mathrsfs}
\usepackage{color}

\begin{document}

\newtheorem{lemma}{Lemma}
\newtheorem{corollary}{Corollary}

\shorttitle{Optimal control of G\"{o}rtler vortices in the nonlinear regime} 
\shortauthor{Sescu \& Afsar} 

\title{Hampering G\"{o}rtler vortices via optimal control in the framework of nonlinear boundary region equations 
}

\author
 {
 Adrian Sescu\aff{1}
  \corresp{\email{sescu@ae.msstate.edu}},
 Mohammed Afsar\aff{2}
  }

\affiliation
{
\aff{1}
Department of Aerospace Engineering, Mississippi State University, Starkville, MS 39762, USA
\aff{2}
Mechanical \& Aerospace Engineering, University of Strathclyde, 16 Richmond St, Glasgow G1 1XQ, UK
}

\maketitle

\begin{abstract}

The control of stream-wise vortices in high Reynolds number boundary layer flows often aims at reducing the vortex energy as a means of mitigating the growth of secondary instabilities, which eventually delay the transition from
laminar to turbulent flow. In this paper, we aim at utilizing such an energy reduction strategy using optimal control theory to limit the growth of G\"{o}rtler vortices developing in an incompressible laminar boundary layer flow over a concave wall, and excited by a row of roughness elements with span-wise separation in the same order of magnitude as the boundary layer thickness. Commensurate with control theory formalism, we transform a constrained optimization problem into an unconstrained one by applying the method of Lagrange multipliers. A high Reynolds number asymptotic framework is utilized, wherein the Navier-Stokes equations are reduced to the boundary region equations (BRE), in which wall deformations enter the problem through an appropriate Prandtl transformation. In the optimal control strategy, the wall displacement or the wall transpiration velocity serve as control variables, while the cost functional is defined in terms of the wall shear stress. Our numerical results indicate, among other things, that the optimal control algorithm is very effective in reducing the amplitude of the G\"{o}rtler vortices, especially for the control based on wall displacement.

\end{abstract}

\section{Introduction}


The control of transitional or fully-developed turbulent boundary layers is intended to reduce the energy carried by stream-wise oriented structures that appear in the form of high- and low-velocity streaks and develop in the near wall region, known to be the starting points of the bursting sequences. Boundary layer streaks over flat plates or wings develop when the height of upstream roughness elements exceeds a certain critical value 
or the amplitude of the freestream disturbances is greater than a given threshold.
Elongated streaks in the form of stream-wise (G\"{o}rtler) vortices also appear inside a boundary layer flow along a concave surface due to imbalance between radial pressure gradients posed by the wall and centrifugal forces. 
From practical standpoint, G\"{o}rtler vortices are important in a number of engineering applications, such as the flow around wings or turbomachinery blades, and the flow in the proximity to the walls of wind tunnels or turbofan engine intakes. The transition from laminar to turbulent boundary layers due to G\"{o}rtler instabilities on the walls of supersonic and hypersonic wind tunnels has been recognized recently as a significant source of noise, which interferes drastically with the measurements in the test section (\cite{Schneider}), inevitably making the comparison between wind tunnel measurements and real flight conditions a challenging problem. Owing to their technological significance, it is desirable to reduce the energy associated with G\"{o}rtler vortices, in an attempt to delay early nonlinear breakdown and the transition into turbulence. Since it is the transient part of the disturbance that dominates the growth of streaks or other three-dimensional disturbances that lead to breakdown, any effective method of control of these streaks must focus on restricting the development of the transient modes. Without claiming to be exhaustive, next subsections will review some of the most important and relevant studies in boundary layer control via wall effects.

\subsection{Control based on wall transpiration}

The control of boundary layers based on wall transpiration can be applied via localized suction regions below low-velocity streaks and blowing regions below high-velocity streaks. The net result is a decrease in the span-wise variation of the stream-wise velocity and, therefore, a commensurate reduction in the number and strength of the bursting events. 

An alternative approach is active wall control, which has been used in the context of turbulent channel flow (see \cite{Choi}) as a means to reduce skin friction drag. \cite{Choi} carried out direct numerical simulations with active wall control based on wall transpiration, by placing sensors in a sectional plane that is parallel to the wall; a frictional drag reduction of approximately 25\% was achieved. From the practical point of view it is difficult (or even impossible) to place sensors in the flow primarily because they may interfere with the disturbances themselves. Therefore, in the same study, \cite{Choi} investigated the same control algorithm but with sensors placed at the wall with information based on the leading term in the Taylor series expansion of the vertical component of velocity near the wall; this approach provided a reduction of only 6\% though. A similar feedback control algorithm was employed by \cite{Koumoutsakos1,Koumoutsakos2}, in which the control is informed again by flow quantities at the wall. A more significant skin friction reduction (approximately 40\%) was obtained by using the vorticity flux components as inputs to the control algorithm. 

\cite{Lee1} derived new suboptimal feedback control laws based on blowing and suction to manipulate the flow structures in the proximity to the wall, using surface pressure or shear stress distribution (the reduction in the frictional drag was in the range of 16-20\%). Observing that the opposition control technique is more effective in low Reynolds number turbulent wall flows, \cite{Pamies} proposed the utilization of the blowing only at high Reynolds numbers, and by doing so they obtained significant reduction in the skin-friction drag for these flows. Recently, \cite{Stroh} conducted a comparison between the opposition control applied in the framework of turbulent channel flow and a spatially developing turbulent boundary layer. They found that the rates of frictional drag reduction are approximately similar in both cases. An overview of the issues and limitations associated with the opposition control type is given in the review article of \cite{Kim1}. 

\cite{Hogberg1} reported the first successful relaminarization of a $Re_{\tau}=100$ turbulent channel flow by applying zero mass flux blowing and suction at the wall in the framework of linear full-state optimal control theory. They showed that the information available in the linearized equations may be sufficient to construct linear controllers able to relaminarize a wall turbulent flow, but this may be limited to low Reynolds number flows.

A number of experiments aiming to control disturbances in laminar or turbulent boundary layers by blowing and suction have been conducted over the years. Several of them are briefly mentioned here. \cite{Gad1} used continuous or intermittent suction to eliminate artificially generated disturbances in a flat-plate boundary layer. The same idea was used in the experiments conducted by Myose and Blackwelder (1995) to delay the breakdown of G\"{o}rtler vortices. \cite{Jacobson} developed a new type of actuation based on a vortex generator to control disturbances generated by a cylinder with the axis normal to the wall, and unsteady boundary layer streaks generated by pulsed suction. Regarding the latter, the actuation was able to significantly reduce the span-wise gradients of the stream-wise velocity, which are known to be an important driving force of secondary instabilities (see \cite{Swearingen}). In the experiments of \cite{Lundell}, stream-wise velocity streaks in a channel flow were controlled by localized regions of suctions in the downstream, which are found to be effective in delaying secondary instabilities and consequently the transition onset. 

\subsection{Control based on wall deformation and motion}

Boundary layer control based on active wall deformations aimed at counteracting streaks in wall turbulence has been successfully applied to reduce the frictional drag, although the reduction was not as high as in the case of opposition control. For this reason, there are no as many studies as there are in the case of blowing and suction. Nevertheless, this type of control has been applied mostly in the framework of turbulent channel flows or turbulent boundary layers, and less in the framework of pre-transitional or transitional boundary layers. 

In one of the earliest studies, performed by \cite{Carlson}, the effect of wall deformation on turbulent structures at the wall was considered; an actuator was used to control one pair of coherent structures near the wall. It was observed that raising the actuator underneath a low-speed streak increases skin-friction drag by allowing the adjacent high-speed region to expand, and vice versa. Another example is the work of \cite{Endo}, reporting DNS studies of feedback control of deformable walls to reduce the skin friction in a turbulent channel flow. The control scheme was based on physical arguments relating to the near-wall coherent structures and a 10\% friction drag reduction was obtained. \cite{Endo} also pointed out that the energy input required to deform the wall is much smaller than the pumping power required for suction/blowing. \cite{Kang} investigated the potential of reducing the skin-friction drag in a turbulent channel flow via active wall motions. They noticed that the instantaneous wall surface shape also took the form of elongated streaks as in laminar boundary layers. A reduction of the friction drag on the order of 13-17\% was realized by their approach. \cite{Mani} utilized a deformable skin actuated by active materials for turbulent boundary layers control, claiming large reduction in skin friction drag. It was based on a generalized actuation principle that is capable of generating a traveling sine wave on the surface of an active skin.

From the experimental standpoint, \cite{Breuer} showed that the energy of nonlinear, non-wave-like disturbances in a boundary layer can be delayed by using a traveling bump at the wall. \cite{Segawa} devised an actuator array to generate wall-normal oscillations, and were able to decrease the regularity of streaky structures (drag reduction was not reported). \cite{Itoh} excited a flexible polythene sheet to generate a transverse travelling wave, which interacted with boundary layer streaks, and achieved 7.5\% drag reduction. Another experimental example is the dissertation research of \cite{Koberg}, where an approach for reducing skin friction in a turbulent boundary layer via active wall deformation was investigated. He attempted to match the velocity sensed away from the wall by imposing a velocity of opposite direction at the wall; the control provided a skin friction reduction of 15\%. \cite{Patzold}, more recently, developed experimentally an actively driven compliant wall to delay the transition in a boundary layer initiated by Tollmien-Schlichting waves. They used various configurations of piezo-actuators combined with an array of sensitive surface sensors, and were able to shift the transition onset forward by 100 mm.

Another relevant control mechanism is the one based on wall oscillations along the stream-wise or span-wise direction. A good portion of the body of research performed in this area is reviewed in \cite{Karniadakis}, and in \cite{Quadrio}. As an example, \cite{Galionis} studied the growth of G\"{o}rtler vortices above a span-wise oscillating surface that is concave in the stream-wise direction, and found a significant reduction in the growth rate associated with secondary instabilities.  Another example is the work of \cite{Hack}, where direct numerical simulations were carried out to study the effect of a span-wise oscillating flat plate on the bypass breakdown to turbulence; they found that the transition onset can be delayed and the transition region can be significantly extended. span-wise wall oscillations have also been shown to attenuate effectively the turbulence intensity in wall-bounded flows, thereby producing a sustained reduction of turbulent wall friction. Experimental (\cite{Laadhari}), numerical (\cite{Quadrio1}) and modelling (\cite{Dhanak}) research studies have appeared since the pioneering study of \cite{Jung}. As an example of early studies, \cite{Choi4} performed an experimental investigation of the effect of span-wise-wall oscillation on the skin friction drag to confirm previous results from numerical simulations. They obtained as high as 45\% drag reduction as a result of some optimizations, and attributed this to the mean velocity gradient reduction due to the span-wise vorticity generated by the Stokes layer. Other studies in this area include \cite{Quadrio2}, \cite{Quadrio3}, \cite{Skote}, \cite{Ricco2}, \cite{Moarref}, \cite{Touber}, \cite{Yakeno}, \cite{Agostini}, or \cite{Hicks}.

Passive wall deformations in the form of riblets ({\it e.g.}, \cite{Walsh}, \cite{Choi3}, \cite{Bechert}, \cite{Lee3}, \cite{Garcia}, \cite{Duan}, \cite{Sasamori}, or \cite{Hou}), compliant surfaces ({\it e.g.}, \cite{Lee4}, \cite{Davies}, \cite{Larose}, \cite{Reutov}, \cite{Carpenter}, \cite{Gad2}), dimples ({\it e.g.}, \cite{Ligrani}, \cite{Wang}, \cite{Lienhart}, or \cite{Tay}), surface waviness ({\it e.g.}, \cite{Du}, \cite{Karniadakis}, \cite{Zverkov}, \cite{Hoepffner}, \cite{Tomiyama}, or \cite{Meysonnat}), or two-dimensional roughness elements ({\it e.g.}, \cite{Holloway}, \cite{Fong}, \cite{Duan2}, or \cite{Park2}) have been applied in a number of studies to delay transition in boundary layers or to reduce the skin friction drag in wall turbulence (these lists of references are far from being comprehensive).

\subsection{Optimal control approach}

Optimal control in the framework of laminar or turbulent boundary layers has been utilized in a number of studies. 
There are numerous studies pertaining the application of optimal control of shear flows (see the review of \cite{Gunzburger} or a more recent review of \cite{Luchini}, although the latter is in a slightly different context). The following studies have targeted the control of disturbances evolving in laminar or turbulent boundary layers ({\it e.g.}, \cite{Bewley}, \cite{Joslin}, \cite{Cathalifaud}, \cite{Corbett}, \cite{Hogberg1}, \cite{Zuccher}, \cite{Cherubini}, \cite{Lu}). 

A relevant work in the present context is that of \cite{Joslin}, where a mathematical framework for optimal control of disturbances in three-dimensional boundary layers based on Lagrange multipliers was introduced; the analysis included in this work is largely based on their derivation. Optimal control of turbulent channel flows by blowing and suction was employed previously by \cite{Bewley}, who claimed a 17\% frictional drag reduction as a result of this scheme. Blowing and suction based optimal control was also applied by \cite{Cathalifaud} to reduce the energy of disturbances in a flat-plate and a concave boundary layer. In the study of \cite{Zuccher}, an optimal and robust control strategy was discussed and tested in the framework of steady three-dimensional disturbances (in the form of streaks) that form in a flat-plate boundary layer. It was based on an adjoint-based optimization technique to first find the optimal state for given initial conditions, and then to determine what the worst initial conditions for the optimal control are. \cite{Lu} derived an optimal control scheme within the linearized unsteady boundary region equations which are the asymptotic reduction of the Navier-Stokes equations under the assumption of low frequency and low stream-wise wavelength. Their study aimed at controlling both streaks developing in flat-plate boundary layers and G\"{o}rtler vortices evolving along concave surfaces. \cite{Cherubini} applied a nonlinear optimal control strategy with blowing and suction, starting with the full Navier-Stokes equations, and using the method of Lagrange multipliers to determine the largest decrease of the disturbance energy. 

A closed-loop optimal control technique based on wall transpiration was derived and tested by \cite{Papadakis}, in the framework of a flat-plate laminar boundary layer excited by freestream disturbances. The optimal control was split into two sequences that can be obtained by marching the corresponding equations in forward and backward directions, and it was found that the feedback sequence is more important than the feed-forward sequence. The study of \cite{Xiao} employs an optimal control algorithm based on Lagrange multipliers, aimed at delaying transition in a flat-plate boundary layer excited by freestream vortical disturbances, is based on blowing and suction, and is derived in the framework of full Navier-Stokes equations.

\subsection{Objectives of this work}

Our aim in this study is to develop an optimal control algorithm that is capable of reducing the stream-wise development of G\"{o}rtler vortices initiated by a row of roughness elements near the leading edge. We show that flow control based on wall deformation or transpiration can significantly reduce the energy of pre-transitional G\"{o}rtler instabilities, which can ultimately delay transition and consequently reduce the skin friction drag. It is shown that the control can be implemented, self-consistently, using the high Reynolds number asymptotic framework for a laminar boundary layer developing along a concave wall, in which the flow is governed by the so-called boundary region equations (BRE) in the nonlinear regime. Since these equations are parabolic in the stream-wise direction, a marching procedure can be utilized to determine the solution based on a given initial/upstream condition. Local changes in the surface geometry are introduced into these equations through a Prandtl transformation (\cite{Yao}) of both dependent and independent variables that does not alter the parabolic character of the boundary value problem and therefore allows solution to be determined by the same numerical marching technique. The variations in local surface shape then allows for a relatively straightforward control strategy to be implemented to the transformed set of equations in order to determine the optimum wall deformation or transpiration that reduces the energy of the vortices. 

The vortices are initiated by perturbing the upstream flow with a periodic array of roughness elements placed near the leading edge using a previously derived asymptotic solution (\cite{Goldstein1,Goldstein2}) in which the appropriate upstream boundary condition was derived. The vortex energy is controlled via an optimal control algorithm based on Lagrange multipliers, where the wall displacement or the wall transpiration velocity serve as control variables. The cost functional is defined in terms of the wall shear stress. In our analysis, local wall deformations appear to resemble elongated surface shapes that we optimize to control the G\"{o}rtler vortex energy. These surface deformations inject/extract momentum into/from the flow in the vertical direction, according to inputs provided by the wall shear stress (the same mechanism is at play when wall transpiration is considered). A similar approach, based on wall deformations, was undertaken in \cite{Sescu2}, where a proportional controller was applied to reduce the energy of G\"{o}rtler vortices. Here, we formulate the problem using the method of Lagrange multipliers, and derive the adjoint BRE equations for control based on both wall deformation and transpiration, and show that significantly more reduction in the G\"{o}rtler vortex energy or wall shear stress can be achieved. It is also shown that the control based on wall deformations is more effective in reducing the streak energy than the control scheme based on wall transpiration, and a potential mechanism behind this difference is discussed.

In section \ref{s2}, the G\"{o}rtler problem is introduced and discussed in the framework of nonlinear boundary region equations, including the basic scalings, associated initial and boundary conditions, and the Prandtl transform utilized to account for wall deformations. Section \ref{s3} introduces the general framework of optimal control, its application to our specific problem, and the derivation of the adjoint equations and optimality conditions (derivation details are given in appendices A, B and C). In section \ref{s4}, various results are presented and discussed for different flow or geometrical conditions, and in section \ref{s5} the physical mechanisms behind both control strategies is discussed. The last section \ref{s6} is reserved for summary and final concluding remarks.

\section{G\"{o}rtler vortices - basic scalings and governing equations}\label{s2}

We consider an incompressible boundary layer flow over a concave surface, with upstream perturbations provided by a span-wise periodic array of roughness elements at some downstream stream-wise location, $x^* = x_0^{*}$ (hereafter all dimensional quantities have a star). The boundary layer flow configuration is the same as that considered in \cite{Sescu1} or \cite{Sescu2} (see figure 1 in \cite{Sescu1}). The effect of the roughness elements are not directly modeled her, but taken into account as initial condition from an asymptotic solution derived previously in \cite{Goldstein1}. The span-wise length scale of the roughness row, $\Lambda^*$, is in the same order of magnitude as the local boundary-layer thickness $\delta^{*} \equiv x_0^{*}/ \sqrt{R} = x_0^{*} \delta$ at the roughness location $x_0^{*}$, where $R = x_0^{*} U_{\infty}^*/\nu^{*}$ is the Reynolds number based on $x_0^{*}$ and $U_{\infty}^*$ is the free stream velocity, with $\nu^{*}$ being the kinematic viscosity, and $\delta \equiv R^{-1/2}$ being the boundary layer thickness scaled by the (fixed) $O(1)$ length scale, $\Lambda^*$. The spatial coordinates are normalized by the span-wise length scale, $\Lambda^*,$ as $(x,y,z)=(x^{*},y^{*},z^{*})/\Lambda^*$, and the velocity and pressure are normalized as

\begin{equation}
\tilde{u}=\frac{u^{*}}{U^{*}_{\infty}},\tilde{v}=R_{\Lambda}\frac{v^{*}}{U^{*}_{\infty}},\tilde{w}=R_{\Lambda}\frac{w^{*}}{U^{*}_{\infty}},\tilde{p}=R_{\Lambda}^{2}\frac{p^{*}}{\rho^{*} U^{*2}_{\infty}}, 
\end{equation}
where $(u^{*},v^{*},w^{*})$ is the dimensional velocity vector, $\rho^{*}$ is the density, and $R_{\Lambda}=U_{\infty}^*\Lambda^*/\nu^*$ is the Reynolds number based on $\Lambda^*$. G\"{o}rtler vortices are expected to develop when $x\sim2\pi/k_{1}\sim\Lambda^* R_{\Lambda}$ (\cite{Wu}), which is fixed at $O(1)$ ($k_1$ is the stream-wise wave number); this suggests the introduction of the slow stream-wise variable $X=x/R_{\Lambda}^*$.
For a body-fitted co-ordinate system, the original Navier Stokes equations are transformed according to the Lam\'{e} coefficients, $h_{1}=(R_{0}-y)/R_{0},h_{2}=1$, where the radius of curvature is much larger than the span-wise separation of the roughness elements; i.e. $R_0 \gg O(1)$. The origin of the coordinate system is located at the leading
edge, with the stream-wise $x$-axis aligned with the wall surface,
$y$-axis perpendicular to the wall surface, and $z$-axis aligned with the
span-wise direction. The velocity field $\{ \tilde{u}, \tilde{v}, \tilde{w} \}$ and the pressure $\tilde{p}$ are expanded like

\begin{equation}\label{asy}
\{ \tilde{u}, \tilde{v}, \tilde{w}, \tilde{p} \} = \{ u(X,y,z), \varepsilon v(X,y,z), \varepsilon w(X,y,z), \varepsilon^2 p(X,y,z) \} + ...
\end{equation}
where the small parameter $\varepsilon = 1/R_{\Lambda}$. 

Upon substituting the dimensionless
independent and dependent variables into the Navier- Stokes equations, with retaining only the first order terms in the asymptotic expansion (\ref{asy}), 
the boundary region equations (BRE) are derived in the form

\begin{equation}\label{nq1}
\frac{\partial u}{\partial X}+\frac{\partial v}{\partial y}+\frac{\partial w}{\partial z}=0,
\end{equation}
\begin{equation}\label{nq2}
u\frac{\partial u}{\partial X}+v\frac{\partial u}{\partial y}+w\frac{\partial u}{\partial z}=\frac{\partial^2 u}{\partial y^2}+\frac{\partial^2 u}{\partial z^2}
\end{equation}
\begin{equation}\label{nq3}
u\frac{\partial v}{\partial X}+v\frac{\partial v}{\partial y}+w\frac{\partial v}{\partial z}+G_{\Lambda}u^{2}=-\frac{\partial p}{\partial y}+\frac{\partial^2 v}{\partial y^2}+\frac{\partial^2 v}{\partial z^2},
\end{equation}
\begin{equation}\label{nq4}
u\frac{\partial w}{\partial X}+v\frac{\partial w}{\partial y}+w\frac{\partial w}{\partial z}=-\frac{\partial p}{\partial z}+\frac{\partial^2 w}{\partial y^2}+\frac{\partial^2 w}{\partial z^2},
\end{equation}
where the effect of the wall curvature is contained in the term involving the global G\"{o}rtler number $G_{\Lambda}=R_{\Lambda}^{2}/R_{0}$ (\cite{Wu}). As discussed in the introduction, the absence of stream-wise second order derivatives in the BRE indicate that they are parabolic in the stream-wise direction and can be solved numerically using a space-marching technique. We will discuss the appropriate boundary conditions below. Prandtl transformation (or Prandtl transposition theorem, \cite{Yao}) is applied to these equations in order to incorporate local changes in wall surface geometry, or wall deformations defined through the function $\mathscr{F}(X,y)$. This is easily done with new wall normal variable and velocity as follows 

\[
Y = y - \mathscr{F}, \hspace{3mm}
\hat{v} = v - \left( u \frac{\partial \mathscr{F}}{\partial X}  + w \frac{\partial \mathscr{F}}{\partial z} \right).
\]
Inserting the chain rule,

\[
 \frac{\partial}{\partial X_i} = 
  \begin{cases}
  \frac{\partial}{\partial X_i} - \frac{\partial \mathscr{F}}{\partial X_i} \frac{\partial }{\partial Y}, i = (1,3),  &  \\
  \frac{\partial }{\partial Y}, i=2,  & 
  \end{cases}
\]
where $\partial/\partial X_i = (\partial/\partial X, \partial/\partial y,\partial/\partial z)$ and $\partial \mathscr{F}/\partial X_i$ is the derivative of the surface function with $X_i = (X,z)$ when $i=(1,3)$, into equations (\ref{nq1})-(\ref{nq4}) gives the transformed BREÕs:

\begin{equation}\label{neq1}
\frac{\partial u}{\partial X}+\frac{\partial \hat{v}}{\partial Y}+\frac{\partial w}{\partial z}=0,
\end{equation}
\begin{equation}\label{neq2}
u\frac{\partial u}{\partial X}+\hat{v}\frac{\partial u}{\partial Y}+w\frac{\partial u}{\partial z}=\frac{\partial^2 u}{\partial Y^2}+\mathscr{D}^2 u - \frac{\partial^2 \mathscr{F}}{\partial z^2} \frac{\partial u}{\partial Y},
\end{equation}
\begin{equation}\label{neq3}
u\frac{\partial v}{\partial X}+\hat{v}\frac{\partial v}{\partial Y}+w\frac{\partial v}{\partial z}+G_{\Lambda}u^{2}=-\frac{\partial p}{\partial Y}+\frac{\partial^2 v}{\partial Y^2}+\mathscr{D}^2 v - \frac{\partial^2 \mathscr{F}}{\partial z^2} v_Y,
\end{equation}
\begin{equation}\label{neq4}
u\frac{\partial w}{\partial X}+\hat{v}\frac{\partial w}{\partial Y}+w\frac{\partial w}{\partial z}=-\mathscr{D} p+\frac{\partial^2 w}{\partial Y^2}+\mathscr{D}^2 w - \frac{\partial^2 \mathscr{F}}{\partial z^2} w_Y,
\end{equation}
where the operator 

\[
\mathscr{D} = \left( \frac{\partial }{\partial z} - \frac{\partial \mathscr{F}}{\partial z} \frac{\partial }{\partial Y} \right)
\]
 was introduced for brevity. These equations obviously retain their parabolic character and are, therefore,
solved by the same marching technique discussed in \cite{Goldstein1} or \cite{Sescu1}. The $v$ variable in (2.8) is treated as an implicit function of $\hat{v}$ via $\hat{v} = v - \left( u \partial \mathscr{F}/\partial X  + w \partial \mathscr{F}/\partial z \right)$ to avoid higher derivatives of $\mathscr{F}$ that might
potentially be difficult to resolve if the wall deformation is a rapidly
varying function of $X$ or $z$ (which is not the case in this particular analysis because the flow variables are assumed to be slowly varying functions of $X$). In this particular study, $\mathscr{F}$ is a continuous and smooth function that will be obtained at discrete points $(X,z)$ within the control algorithm, where $\mathscr{F}=0$ corresponds to the original undeformed surface. The wall displacement and the height of the roughness elements are assumed to be in the order of magnitude of, or smaller than, the boundary layer displacement thickness; otherwise, the theory will fail to provide accurate results.

The wall boundary condition for the control based on wall deformation (as described in the next section) is the no-slip condition

\[
u(X,Y,z)=v(X,Y,z)=w(X,Y,z)=0,
\]
where the shape of the wall is embedded in the transformed wall-normal variable $Y$, that now represents level surfaces where $y(X,z) = \mathscr{F}(X,z)=const$. The wall boundary condition for the control based on wall transpiration is

\[
u(X,Y,z)=w(X,Y,z)=0, v(X,Y,z)=v_w(X,z),
\]
where $v_w(X,z)$ is the transpiration velocity.

\begin{figure}
 \begin{center}
   \includegraphics[width=9cm]{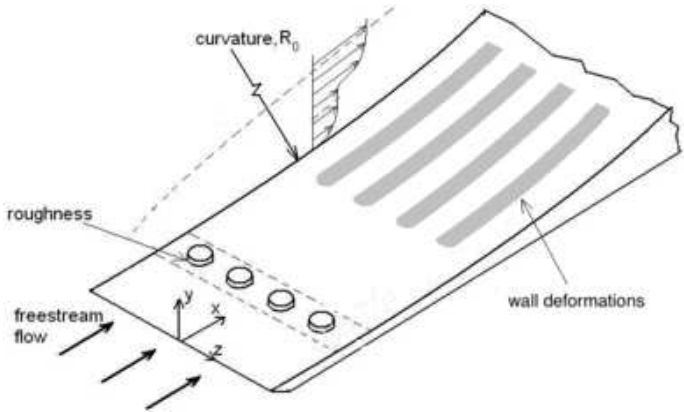}
 \end{center}
  \caption{ Flow configuration sketch.}
  \label{f1}
\end{figure}

The vortices are initiated by roughness elements placed close to the leading edge in which the local surface is geometrically flat owing to the requirement that $R_0 \gg O(1)$ (see figure \ref{f1}). Hence the initial conditions for the transformed BREÕs are given by the flat plate case solution as derived in \cite{Goldstein1} (see also the appendix of \cite{Sescu1}). Equations (\ref{neq1})-(\ref{neq4}) and the associated initial and boundary conditions are marched in the stream-wise direction since they are parabolic. To avoid decoupling between the pressure and velocity, a staggered grid is employed in the wall normal direction, and second-order accurate difference schemes are employed along both $y$ and $z$ directions. The convergence was achieved by a relaxation method, similar to the one employed in \cite{Sescu1}.

\section{Optimal control problem in the nonlinear regime}\label{s3}


While it is common for an optimal flow control problem to be formulated in the framework of a dynamical system (usually, described by a set of equations that are parabolic in time), here we replace the time direction by the $X$-direction owing to the parabolic character of the BRE in the stream-wise direction, which necessitates stream-wise marching to obtain a solution. To fix ideas, we first write equations (\ref{neq1})-(\ref{neq4}) in the generic and more compact form

\begin{equation}\label{oo}
\mathscr{G}(\textbf{q},\psi) = 0,
\end{equation}
for brevity, with initial and boundary conditions

\begin{equation}\label{}
\textbf{q}(0,Y,z) = \textbf{q}_0(Y,z)
\end{equation}
\begin{equation}\label{}
\textbf{q}(X,0,z) = \phi, \hspace{3mm}  \lim_{Y \rightarrow \infty} \textbf{q}(X,Y,z) = \textbf{q}_B(X,z),
\end{equation}
along the wall-normal direction, and periodic or symmetry boundary conditions in the span-wise direction, $z$. In equation (\ref{oo}),
$\mathscr{G}()$ is the non-linear BRE differential operator in abstract notation, $\textbf{q} = (u,v,w,p)$ is the vector of state variables, $\psi$ is the control variable or design parameter that is part of the state equations (in the present case it represents  the functional $\mathscr{F}(X,z)$ describing the wall deformation), $\phi$ is the control variable associated with the boundary conditions ({\it e.g.}, the transpiration velocity at the wall, $v_w$), $\textbf{q}_0(Y,z)$ represents the upstream or initial condition in $X=0$, and $\textbf{q}_B$ is a given function that specifies the boundary condition at infinity. We define an objective (or cost) functional as

\begin{equation}\label{zz}
\mathscr{J}(\textbf{q},\psi,\phi) = \mathscr{E}(\textbf{q}) 
 +  \sigma_1 \left(\| \psi_X \|^{\beta_1} + \| \psi \|^{\beta_1} \right) 
 + \sigma_2 \left(\| \phi_X \|^{\beta_2} + \| \phi \|^{\beta_2} \right),
\end{equation}
where $\mathscr{E}(\textbf{q})$ is a specified target function to be minimized ({\it e.g.}, the energy of the disturbance, or the wall shear stress; the latter is considered in this study), the second and the third terms on the right hand side of (\ref{zz}) are penalization terms depending on the norm of the control variable (usually, these quantities place constraints on the magnitude of the admissible control variables, since they cannot increase or decrease indefinitely), $\sigma_i$ and $\beta_i$, $i=1,2$, are given constants, and subscript $X$ denotes derivative with respect to $X$. The norm $\| \|$ in equation (\ref{zz}) is associated with an appropriate inner product of two complex functions, $f$ and $g$, defined as 

\begin{equation}\label{aaa}
\langle f, g \rangle = \int_{0}^{X_t} f^* g dX
\end{equation}
in the space $[0,X_t]$, with $X_t$ being the terminal stream-wise location (the star in (\ref{aaa}) denotes complex conjugate).

A common approach to transform a (nonlinear) constrained optimization problem into an unconstrained problem is by using the method of Lagrange multipliers (see, for example, \cite{Joslin}, \cite{Gunzburger}, \cite{Zuccher}). To this end, we consider the Lagrangian

\begin{equation}\label{}
\mathscr{L}(\textbf{q},\psi,\phi,\textbf{q}^a) = \mathscr{J}(\textbf{q},\psi,\phi) - \langle \mathscr{G}(\textbf{q},\psi), \textbf{q}^a \rangle,
\end{equation}
where $\textbf{q}^a$ is the vector of Lagrange multipliers $(u^a,v^a,w^a,p^a)$, also known as the adjoint vector. In other words, the Lagrange multipliers are introduced in order to transform the minimization of $\mathscr{J}(\textbf{q},\psi,\phi)$ under the constraint $\mathscr{G}(\textbf{q},\psi) = 0$ into the unconstrained minimization of $\mathscr{L}(\textbf{q},\psi,\phi,\textbf{q}^a)$. The unconstrained optimization problem can be formulated as: 

{\it Find the control variables $\psi$ and $\phi$, the state variables $\textbf{q}$, and the adjoint variables $\textbf{q}^a$ such that the Lagrangian $\mathscr{L}(\textbf{q},\psi,\phi,\textbf{q}^a)$ is a stationary function, that is}

\begin{equation}\label{ab}
\delta \mathscr{L} = \frac{\partial \mathscr{L}}{\partial \textbf{q}} \delta \textbf{q}
                             + \frac{\partial \mathscr{L}}{\partial \psi} \delta \psi
                             + \frac{\partial \mathscr{L}}{\partial \phi} \delta \phi
                             + \frac{\partial \mathscr{L}}{\partial \textbf{q}^a} \delta \textbf{q}^a = 0
\end{equation}
where 

\begin{equation}\label{}
\frac{\partial \mathscr{L}}{\partial a} \delta a = \frac{\mathscr{L}(a+\epsilon \delta a) - \mathscr{L}(a)}{\epsilon}
\end{equation}
represents directional differentiation in the generic direction $\delta a$. All directional derivatives in (\ref{ab}) must vanish, providing different sets of equations: 

\begin{itemize}

\item adjoint BRE equations are obtained by taking the derivative with respect to $\textbf{q}$,

\begin{eqnarray}\label{ii1}
\frac{\partial \mathscr{L}}{\partial \textbf{q}} = 0  \hspace{4mm} \Rightarrow   \hspace{4mm}   \mathscr{G}^a(\textbf{q}^a,\psi) = 0  
\end{eqnarray}

\item optimality conditions are obtained by taking the derivatives with respect to $\psi$ or $\phi$,

\begin{eqnarray}\label{ii2}
\frac{\partial \mathscr{L}}{\partial \psi} = 0  \hspace{2mm},  \hspace{2mm} \frac{\partial \mathscr{L}}{\partial \phi} = 0  \hspace{4mm} \Rightarrow   \hspace{4mm}   \mathscr{O}(\textbf{q}^a,\textbf{q},\psi,\phi) = 0 
\end{eqnarray}

\item the original BRE equations are obtained by taking the derivative with respect to $\textbf{q}^a$,

\begin{eqnarray}\label{ii3}
\frac{\partial \mathscr{L}}{\partial \textbf{q}_a} = 0  \hspace{4mm} \Rightarrow   \hspace{4mm}   \mathscr{G}(\textbf{q},\psi) = 0.
\end{eqnarray}

\end{itemize}

Equations (\ref{ii1})-(\ref{ii3}) form the optimal control system that can be utilized to determine the optimal states and the control variables. One can note that the stationarity of the Lagrangian with respect to the adjoint variables $\textbf{q}^a = (u^a,v^a,w^a,p^a)$ essentially yields the original state equations, while the stationarity with respect to the state variables $\textbf{q} = (u,v,w,p)$ yields the adjoint equations that depend on the state variables. The relationship between the state variables and the adjoint variables can be expressed by the adjoint identity,

\begin{eqnarray}\label{}
\langle \mathscr{G}(\textbf{q},\psi), \textbf{q}^a \rangle = \langle \textbf{q}, \mathscr{G}^a(\textbf{q}^a,\psi) \rangle 
+ \mathscr{B}(\phi)
\end{eqnarray}
where the last term, $\mathscr{B}$, represents a residual from the boundary conditions.

The adjoint BRE equations (see derivation in appendix A) are

\begin{equation}\label{adj1}
\frac{\partial v^a}{\partial Y} + \mathscr{D} w^a = 0,
\end{equation}
\begin{eqnarray}\label{adj2}
&-& u\frac{\partial u^a}{\partial X} - \hat{v}\frac{\partial u^a}{\partial Y} - w\frac{\partial u^a}{\partial z} + u^a \frac{\partial u}{\partial X}+v^a \frac{\partial v}{\partial X}+w^a \frac{\partial w}{\partial X} -  \frac{\partial \mathscr{F}}{\partial X} (u^a \frac{\partial u}{\partial Y}+v^a \frac{\partial v}{\partial Y}+w^a \frac{\partial w}{\partial Y})   \nonumber \\
&+& 2 G_{\Lambda}u v^a
- \frac{\partial p^a}{\partial X} + \frac{\partial \mathscr{F}}{\partial X} \frac{\partial p^a}{\partial Y} - \frac{\partial^2 u^a}{\partial Y^2} - \mathscr{D}^2 u^a - \frac{\partial^2 \mathscr{F}}{\partial z^2} \frac{\partial u^a}{\partial Y} = 0,
\end{eqnarray}
\begin{eqnarray}\label{adj3}
&-& u\frac{\partial v^a}{\partial X} - \hat{v}\frac{\partial v^a}{\partial Y} - w\frac{\partial v^a}{\partial z} + u^a \frac{\partial u}{\partial Y}+v^a \frac{\partial v}{\partial Y}+w^a \frac{\partial w}{\partial Y} - \frac{\partial p^a}{\partial Y} - \frac{\partial^2 v^a}{\partial Y^2} \nonumber \\ 
&-& \mathscr{D}^2 v^a - \frac{\partial^2 \mathscr{F}}{\partial z^2} \frac{\partial v^a}{\partial Y} = 0,
\end{eqnarray}
\begin{eqnarray}\label{adj4}
&-& u\frac{\partial w^a}{\partial X} - \hat{v}\frac{\partial w^a}{\partial Y} - w\frac{\partial w^a}{\partial z} + u^a \frac{\partial u}{\partial z}+v^a \frac{\partial v}{\partial z}+w^a \frac{\partial w}{\partial z} -  \frac{\partial \mathscr{F}}{\partial z} (u^a \frac{\partial u}{\partial Y}+v^a \frac{\partial v}{\partial Y}+w^a \frac{\partial w}{\partial Y})   \nonumber \\
&-& \mathscr{D} p^a - \frac{\partial^2 w^a}{\partial Y^2} - \mathscr{D}^2 w^a - \frac{\partial^2 \mathscr{F}}{\partial z^2} \frac{\partial w^a}{\partial Y} = 0,
\end{eqnarray}
satisfying the initial and boundary conditions

\begin{equation}\label{bc1}
(u^a,v^a,w^a,p^a)|_{X=X_t} = (0,0,0,0) \hspace{2mm} {\textnormal i\textnormal n}  \hspace{2mm} \Omega,
\end{equation}
\begin{equation}\label{bc2}
 (u^a,v^a,w^a)|_{\Gamma} = 
  \begin{cases}
  (\alpha(\tau_w - \tau_0),0,0) \hspace{2mm} \textnormal f\textnormal o\textnormal r  \hspace{2mm} X \in [X_{s0},X_{s1}] &  \\
  (0,0,0) \hspace{2mm} \textnormal o\textnormal t\textnormal h\textnormal e\textnormal r\textnormal w\textnormal i\textnormal s\textnormal e  & 
  \end{cases}
\end{equation}
and 

\begin{equation}\label{bc3}
(u^a,v^a,w^a,p^a)|_{Y \rightarrow \infty} = (0,0,0,0) \hspace{2mm}
\end{equation}
where $\alpha$ is a constant pre-factor that controls the penalization of the wall shear stress. With the state variables $(u,v,w,p)$ determined from equations (\ref{neq1})-(\ref{neq4}), the adjoint equations (\ref{adj1})-(\ref{adj4}) are linear and parabolic, and can be solved via a marching procedure in the backward direction, starting from the terminal stream-wise location, $X_t$, towards the initial stream-wise location $X_0$. 

The optimality equation for control based on wall transpiration is obtained in the form (see derivation in Appendix B)

\begin{equation}\label{nn1}
\frac{\partial^2 v_w}{\partial X^2} - v_w = \frac{1}{\sigma_2} \left( p^a + \frac{\partial v^a}{\partial Y}  \right),
\end{equation}
satisfying the boundary conditions

\begin{equation}\label{mm1}
\frac{\partial v_w}{\partial X}(X_0) = 0, \frac{\partial v_w}{\partial X}(X_1) = 0,
\end{equation}
Given $p^a$ and $v^a$ from the adjoint equations, (\ref{nn1}) represents a two-point boundary value problem for $X$ inside the interval $[X_0,X_1]$. The optimality equation for control based on wall deformation is given as (see derivation in Appendix C)

\begin{eqnarray}\label{nn2}
 \sigma_1 \frac{\partial^2 \mathscr{F}}{\partial X^2} - 2 \frac{\partial u^a}{\partial z} \frac{\partial^2 u}{\partial Y^2}  \frac{\partial \mathscr{F}}{\partial z}  - \sigma_1 \mathscr{F}
= \frac{\partial^2 u^a}{\partial z^2} \frac{\partial u}{\partial Y}- 2 \frac{\partial u^a}{\partial z} \frac{\partial^2 u}{\partial Y\partial z}  - \frac{\partial p^a}{\partial X} \frac{\partial u}{\partial Y} - \frac{\partial p^a}{\partial z} \frac{\partial w}{\partial Y},
\end{eqnarray}
subject to the boundary conditions

\begin{equation}\label{mm2}
\frac{\partial \mathscr{F}}{\partial X}(X_0) = 0, \frac{\partial \mathscr{F}}{\partial X}(X_1) = 0.
\end{equation}
Again, (\ref{nn2}) represents a two-point boundary value problem for X inside the interval $[X_0,X_1]$, but unlike the optimality condition associated with the wall transpiration that depends only on adjoint variables, this equation also depends on state variables, $u$ and $w$, and their derivatives. The constant pre-factors $\sigma_1$ and $\sigma_2$ are used to control the importance of the terms they multiply in (\ref{opt}).

Alternatively, the optimality condition (\ref{nn1}) or (\ref{nn2}) can be replaced by a gradient method or steepest descent method, wherein the objective function is updated according to

\begin{equation}\label{}
\mathscr{F}^{(n+1)} = \mathscr{F}^{(n)} - \alpha \frac{d \mathscr{J}^{(n)}} {d \mathscr{F}^{(n)}}
\end{equation}
for control based on wall deformations, or

\begin{equation}\label{}
v_w^{(n+1)} = v_w^{(n)} - \alpha \frac{d \mathscr{J}^{(n)}} {d v_w^{(n)}}
\end{equation}
for control based on wall transpiration, where $n$ represent the iteration index, and $\alpha$ in this case is the descent parameter (note that the results in this study are obtained based on solving the optimality equations (\ref{nn1}) and (\ref{nn2})).

\begin{figure}
 \begin{center}
   \includegraphics[width=12cm]{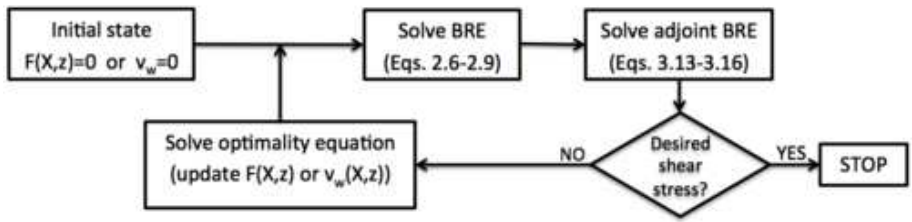}
 \end{center}
  \caption{ Schematic of the control algorithm steps.}
  \label{f2}
\end{figure}

The optimal control procedure based on wall transpiration or deformation is shown schematically in figure \ref{f2}. The control algorithm starts with the solution to BRE for the uncontrolled boundary layer, followed by the solution to the adjoint BRE (note that the adjoint BRE depend on the solution to the BRE). The difference between the wall shear stress and the Blasius wall shear stress is then compared to a desired value; if the difference is larger than a given threshold then the optimality condition equation is solved to determine the new wall shape $\mathscr{F}$, or the new wall transpiration velocity $v_w$. Once these are determined, the loop is repeated, and the calculation finishes when the difference between the wall shear stress and the desired value is less than the threshold.

Adjoint equations (\ref{adj1})-(\ref{adj4}) and the associated initial and boundary conditions (\ref{bc1})-(\ref{bc3}) are solved numerically on the same grid as the original BRE state equations (\ref{neq1})-(\ref{neq4}), and utilizing the same numerical algorithm (as for the BRE equations), except that the marching is preformed backwards, starting from the terminal stream-wise location. The optimality equations (\ref{nn1}) and (\ref{nn2}) and the associated boundary conditions (\ref{mm1}) and (\ref{mm2}) respectively, are solved via a Jacobi relaxation method.

\section{Results and discussion}\label{s4}

\subsection{Preliminaries}

A row of roughness elements located at a fixed distance of $0.5$ m from the leading edge (see figure \ref{f1}) is considered for the span-wise separation distances $\Lambda^*=0.8$, $1.2$ and $1.6$ cm. The effect of roughness elements on the boundary layer is not directly modeled, but taken into account by imposing initial conditions derived in \cite{Goldstein1}through asymptotic analysis under the assumption that the roughness height is asymptotically small and the Reynolds number is large. The G\"{o}rtler number based on span-wise separation is $G_{\Lambda}=318887$ (corresponding to a G\"{o}rtler number based on the momentum displacement thickness of $6.428$). The functional form describing the roughness element is given by the localized function
$
exp[-(x^2 + z^2)/ (D/2)^2 ],
$
where $D$ represents the diameter of the roughness element. The kinetic energy associated with the G\"{o}rtler vortex is calculated according to

\begin{equation}\label{jj}
E(X) = \intop_{z_1}^{z_2}  \intop_{0}^{\infty}  \left[ \left| v - v_m(X,y) \right|^{2} +  \left| w - w_m(X,y) \right|^{2} +  \left| u - u_m(X,y) \right|^{2} \right] dzdy,
\end{equation}
where $u_m(X,y)$, $v_m(X,y)$, and $w_m(X,y)$ are the span-wise mean components of velocity, and $z_1$ and $z_2$ are the boundary limits in the span-wise direction, while the span-wise averaged wall shear stress given by

\begin{equation}\label{jj}
\tau_w(X) = \frac{1}{(z_2 - z_1)}\intop_{z_1}^{z_2}    \frac{\partial u}{\partial Y}(X,0,z)  dz.
\end{equation}
where, as mentioned earlier, the shape of the wall is taken into account implicitly through the partial derivative with respect to $Y$.

In the results that follow, control is applied using either wall deformation or wall transpiration by an iterative procedure as described in the previous section. In the case of wall transpiration, the cumulative blowing or suction at the wall is realized such that a zero mass flow rate is maintained to avoid injecting or absorbing mass into/from the flow. The efficiency of the control is quantified through results consisting of stream-wise velocity contours, stream-wise distribution of the energy of the disturbances and of the span-wise averaged wall shear stress.

\begin{figure}
 \begin{center}
   \includegraphics[width=5.5cm]{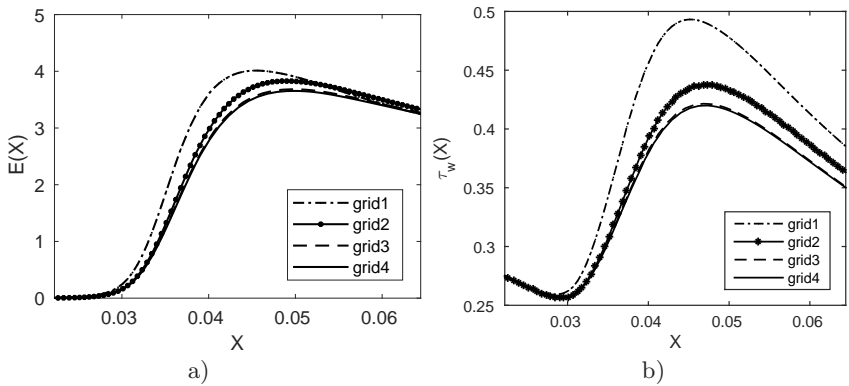} 
   \includegraphics[width=5.5cm]{grid_study_shear} \\
a)  \hspace{55mm}  b) \\
 \end{center}
  \caption{ Energy (left) and the span-wise averaged wall shear stress (right) calculated using four different grids.}
  \label{f3}
\end{figure}

To ensure that the numerical results are accurate, a grid study is performed to determine the appropriate number of grid points along the wall-normal and span-wise directions. To this end, we consider four grid resolutions, consisting of $21$, $31$, $41$ and $51$ points in the span-wise direction, and $141$, $161$, $181$ and $201$ points in the wall-normal direction, while the number of grid points in the stream-wise direction was fixed at $300$, corresponding to a spatial step of $\Delta X = 2.222\times 10^{-4}$. To reduce the computational cost, the domain size in the span-wise direction has been reduced to half distance between two roughness elements, with symmetry conditions - instead of periodic conditions - applied at the right and left boundaries. Results in terms of energy of the disturbance and span-wise averaged shear stress distributions in the stream-wise direction are plotted in figure \ref{f3} for all four grids. The solution converges to a final state as the number of grid points is increased (the convergence is measured by the relative energy error between two iterations, which was set at $10^{-3}$). As a result of the grid study, the resolution associated with 'grid3', consisting of $41\times181$ points in the span-wise and wall-normal directions respectively, was chosen for subsequent calculations. The numerical solution was found to be less sensitive to the resolution in the stream-wise direction because a sufficiently large number of grid points was chosen (appendix D shows a comparison between three resolutions).

\begin{figure}
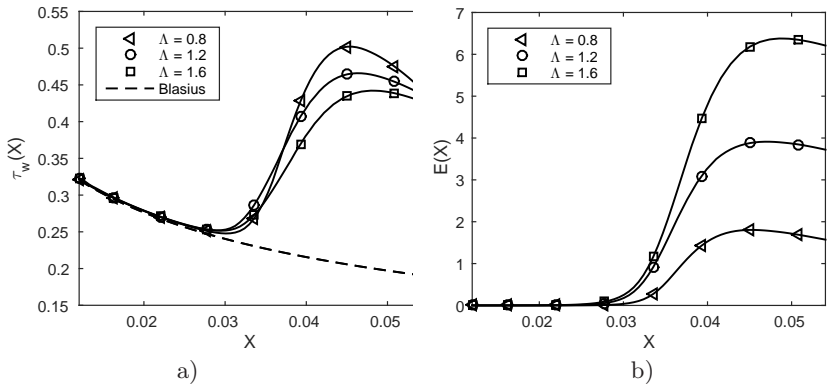

 \begin{center}
   \includegraphics[width=5.5cm]{shear_lam} 
   \includegraphics[width=5.25cm]{ener_lam} \\
a)  \hspace{55mm}  b) \\
 \end{center}
  \caption{ Span-wise averaged wall shear stress (left) and energy (right) as a function of the stream-wise direction and for different roughness span-wise separation.}
  \label{f4}
\end{figure}

Before presenting results from the control schemes, in figure \ref{f4}, we show the effect of varying the span-wise separation on the vortex energy growth (left) and on the span-wise averaged wall shear stress (right) for the uncontrolled boundary layer. It is seen that while the wall shear stress is not substantially affected, the energy of the disturbance seems to show a significant variation.

\begin{figure}
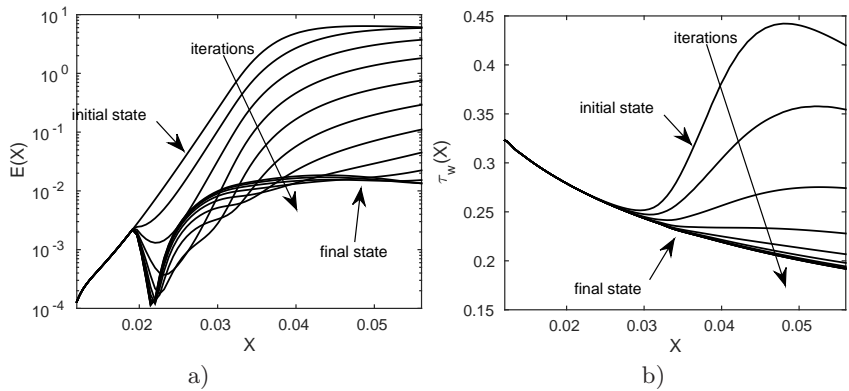

 \begin{center}
   \includegraphics[width=5.5cm]{energy_convergence} 
   \includegraphics[width=5.5cm]{shear_convergence} \\
a)  \hspace{55mm}  b) \\
 \end{center}
  \caption{ Typical convergence of the control algorithm: a) energy; c) span-wise averaged wall shear stress.}
  \label{f5}
\end{figure}

In the following subsections, the interval where the control is applied, $[X_0,X_1]$, and the interval where the cost function is defined, $[X_{s0},X_{s1}]$, are the same; the downstream boundary of both intervals coincides with the terminal location $X_t$. To illustrate the control iterations and its convergence, figure \ref{f5} shows the energy of the disturbance (left) and the span-wise averaged wall shear stress (right) for a control algorithm based on wall deformations; for this particular case, about $12$ iterations were necessary to obtained the optimum wall deformation that provides the lowest wall shear stress (the total number of iterations depends on the parameters $\sigma_1$ and $\sigma_2$ in equation (\ref{zz})). The boundary layer was excited by a row of periodic roughness elements (the roughness elements are located in $x^* = 0.5$ m from the leading edge, the span-wise separation is $1.2$ cm, and $X_0=0.015$). The initial state curve in figure \ref{f5} corresponds to the energy associated with the uncontrolled G\"{o}rtler vortices, while the final state curve represents the optimum wall deformation. More than three orders of magnitude reduction in the energy was attained for this particular case. An exponential convergence for the first several steps can be observed in figure \ref{f5}a, where the energy increment in the logarithmic scale from one iteration to the other seems to be constant. The shear stress plot on the right (part b) shows that in 5-6 iterations the level of the frictional drag can be reduced considerably (the final state is remarkably very close to the shear stress corresponding to the original Blasius solution as it will be shown in the subsequent figures).

\subsection{Sensitivity to control location, $X_0$}

\begin{figure}
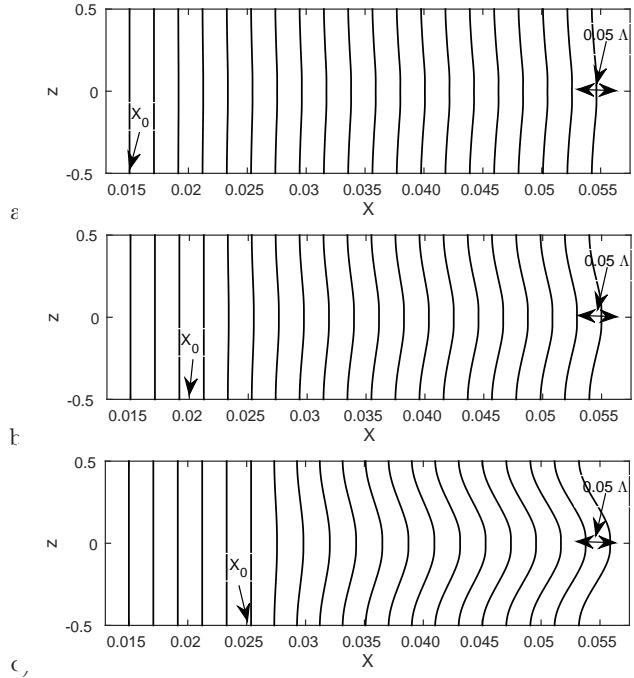

 \begin{center}
 a)  \includegraphics[width=8cm]{surf_wall_lam12} \\
 b)  \includegraphics[width=8cm]{surf_wall_lam12_1} \\
 c)  \includegraphics[width=8cm]{surf_wall_lam12_2} \\
 \end{center}
  \caption{ The displacement of the wall represented by profiles in the z-direction, for the span-wise separation $\Lambda^* = 1.2$ and different control initiation locations: a) $X_0 = 0.015$; b) $X_0 = 0.02$; c)  $X_0 = 0.025$.}
  \label{f6}
\end{figure}

In this section, we show results for the optimal control applied for three different values of $X_0$, which corresponds to the location where the control is initiated. Both types of control, wall deformation and transpiration, are analyzed here, and results for span-wise separation of $\Lambda^*=1.2$ are presented and discussed. The wall displacement as a consequence of the control based on wall deformations is plotted in figure \ref{f6} at different stream-wise locations, and for all three values of $X_0$. They represent distributions of the wall displacement along the span-wise direction at equal stream-wise increments starting from $X=0.015$ to $X=0.055$. It appears that the amplitude of the wall displacement is highly dependent on $X_0$, and that the displacement is a small percent from the span-wise separation (the maximum is approximately $5\%$, which corresponds to the largest value of $X_0$). For the smallest value of $X_0$, the maximum wall displacement is below $1\%$, which can be considered negligible in the calculation of the integrated frictional drag (i.e., there is no considerable increase in the surface area due to the wall displacement). These results suggest that as the energy of the G\"{o}rtler vortex increases with the stream-wise coordinate it becomes more difficult to control the flow using wall deformations.

\begin{figure}
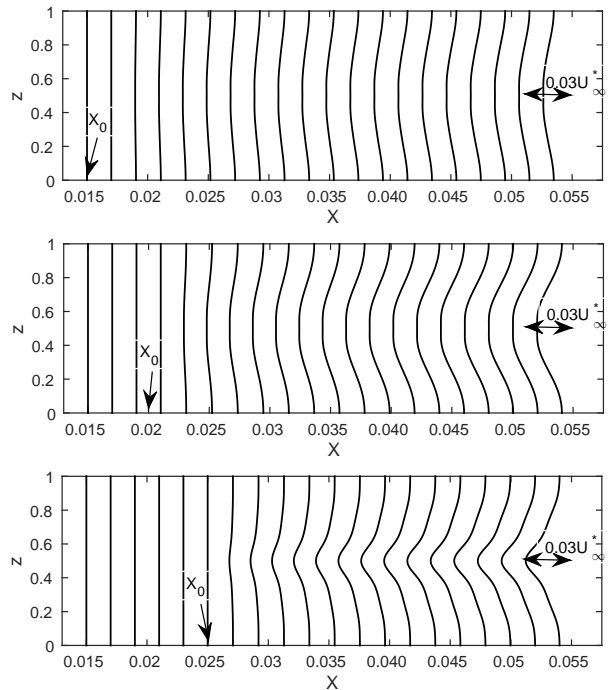

 \begin{center}
 a)  \includegraphics[width=8cm]{surf_blo_lam12} \\
 b)  \includegraphics[width=8cm]{surf_blo_lam12_1} \\
 c)  \includegraphics[width=8cm]{surf_blo_lam12_2} \\
 \end{center}
  \caption{ The distribution of transpiration velocity, $v_w/V_{\infty}$, at the wall represented by profiles in the span-wise z-direction, for the span-wise separation $\Lambda^* = 1.2$ and different control initiation locations: a) $X_0 = 0.015$; b) $X_0 = 0.02$; c)  $X_0 = 0.025$.}
  \label{f7}
\end{figure}

Similar distribution of curves for different values of $X_0$ are included in figure \ref{f7}, except the wall transpiration velocity is plotted along the span-wise direction. As the control initiation location is moved downstream the magnitude of the transpiration velocity is increased, which is to be expected because the streak strength (defined as the difference between the high speed and the low speed in the streaks) increases in the stream-wise direction, demanding for more momentum injection or suction from the wall. It will be shown in the next contour plots that moving the initiation location in the downstream is not as effective in reducing the energy of the vortices as the control based on wall deformations. The maximum transpiration velocity attained in the downstream for $X_0=0.025$ is a small percentage from the freestream velocity (3\%); it is a much smaller percent for $X_0=0.015$ (less than 1\%).

\begin{figure}
 \begin{center}
      \includegraphics[width=4.4cm]{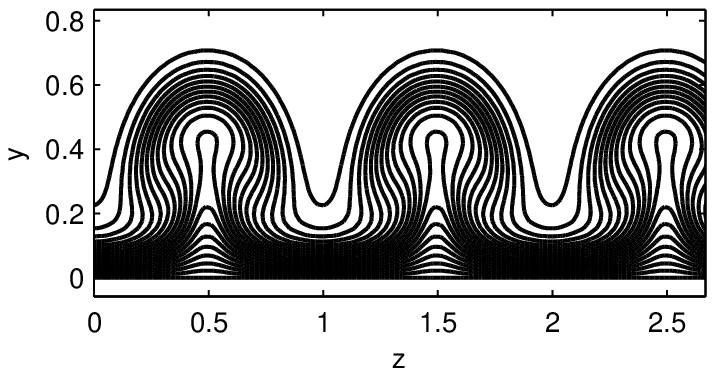} \\
\hspace{5mm}  a)  \\
      \includegraphics[width=4.4cm]{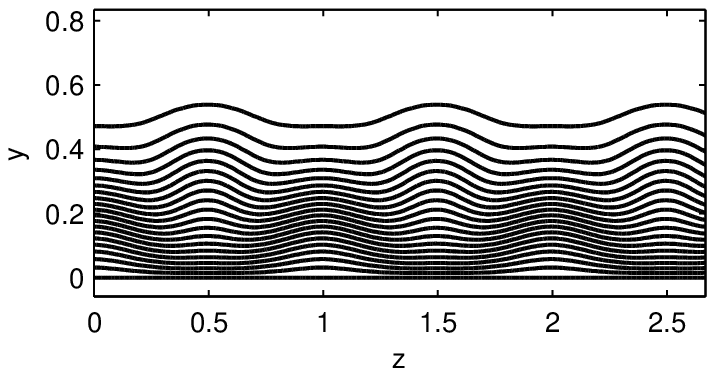}
      \includegraphics[width=4.4cm]{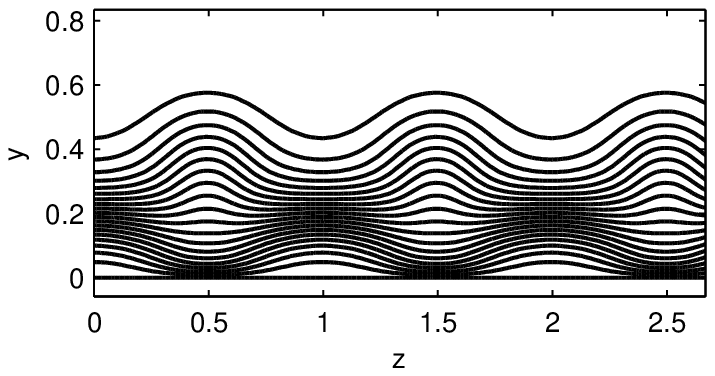}
      \includegraphics[width=4.4cm]{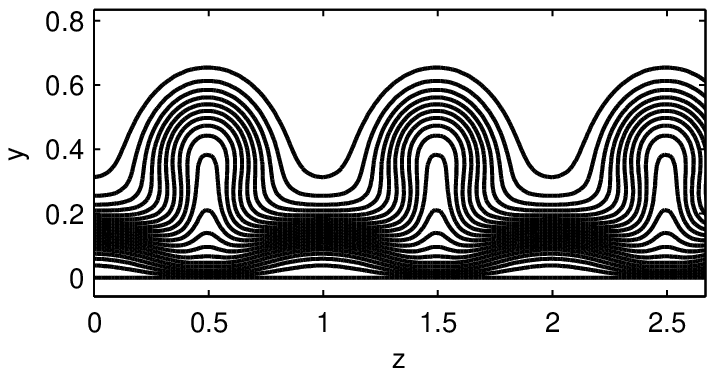} \\
\hspace{5mm}  b)  \hspace{40mm}  c)  \hspace{40mm}  d) \\
      \includegraphics[width=4.4cm]{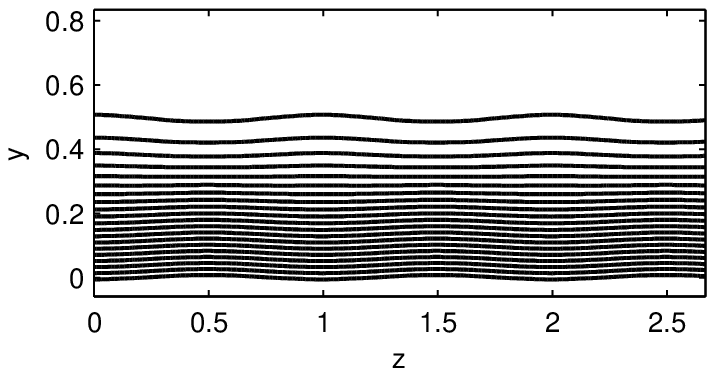}
      \includegraphics[width=4.4cm]{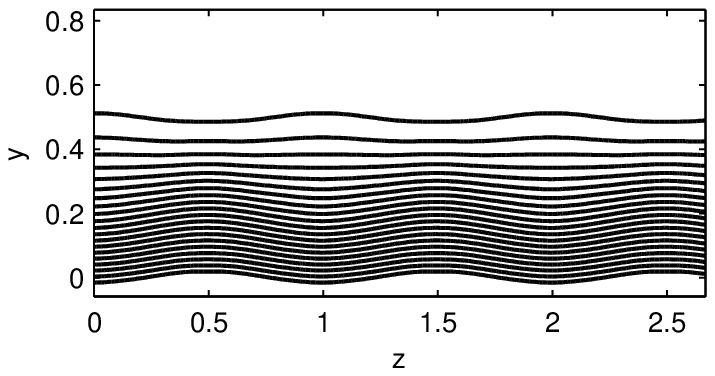}
      \includegraphics[width=4.4cm]{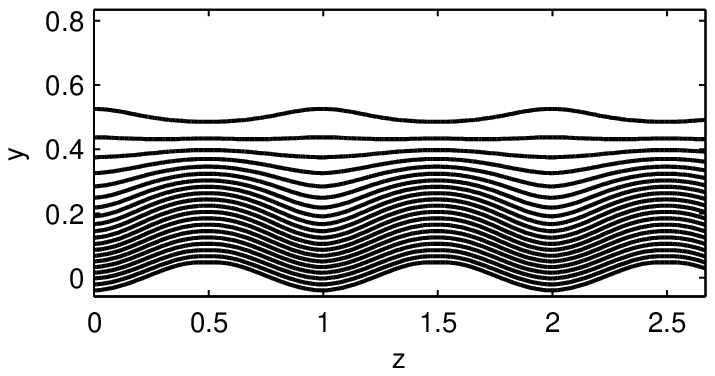} \\
\hspace{5mm}  e)  \hspace{40mm}  f)  \hspace{40mm}  g) \\
 \end{center}
  \caption{ Stream-wise velocity contours for span-wise separation of $1.2$ cm, at the stream-wise location $X = 0.04$: a) no control; b) control based on wall transpiration, $X_0 = 0.015$; c) control based on wall transpiration, $X_0 = 0.02$; d) control based on wall transpiration, $X_0 = 0.025$; e) control based on wall deformation, $X_0 = 0.015$; f) control based on wall deformation, $X_0 = 0.02$; g) control based on wall deformation, $X_0 = 0.025$.}
  \label{f8}
\end{figure}

Regions of low- and high-speed associated with the streak can be visualized by plotting contours of stream-wise velocity in sections at the fixed stream-wise location $X = 0.04$, as shown in figure \ref{f8}. G\"{o}rtler vortices developing in the uncontrolled case (figure \ref{f8}a) reveal fully-developed `mushroom` shapes with alternating low-speed streaks and high-speed streaks in the span-wise direction. Parts b, c and d of figure \ref{f8} correspond to results obtained by applying the control based on wall transpiration, while the right column of plots (parts e, f and g) corresponds to results obtained by applying wall deformations. In the control algorithm based on wall deformations, the wall surface is gradually moved upward at the span-wise location corresponding to the low-speed streak, while at the same time it is moved downward at the span-wise location corresponding to high-speed streaks. This change in the geometry of the wall surface increases or decreases the momentum of the flow, thus reducing the energy associated with the G\"{o}rtler vortices. The same mechanism is at play when wall transpiration is applied, except that the upward motion of the wall is replaced by blowing while the downward motion by suction. These contours show that the control based on wall deformations is more effective than the control based on wall transpiration, especially at high control initiation location $X_0=0.025$. The difference between the results from the two control schemes is qualitatively measured from the contour plots by comparing the level of span-wise variation; for example, by comparing figures \ref{f8}b and \ref{f8}c it is noted that there are significant span-wise variations of the contour lines on the left hand side (corresponding to the control based on wall transpiration), while on the right hand side the flow appears as a Blasius boundary layer; more quantitative results will be shown in plots of energy and wall shear stress later in the section. Note that previously, it was found that the control based on wall transpiration is more effective than control based on wall deformation in wall turbulence (see, for example, \cite{Endo}, \cite{Kang}). However, the comparison of our results with the control of wall turbulence is largely irrelevant since the physics of the flow is different; moreover, here we target steady G\"{o}rtler vortices, where the stream-wise gradients are neglected, while a turbulent boundary layer is highly unsteady with strong variations of the flow in the stream-wise direction. Another reason is that for our problem a parabolic set of equations is solved in a pre-transitional boundary layer, where there is no feedback to the downstream flow, and the variation of the wall deformation with respect to the stream-wise direction is rather gentle (versus the variation in the span-wise direction). A thorough discussion of the physical mechanism and interpretations behind both types of control algorithms is presented later in section \ref{s5}.

\begin{figure}
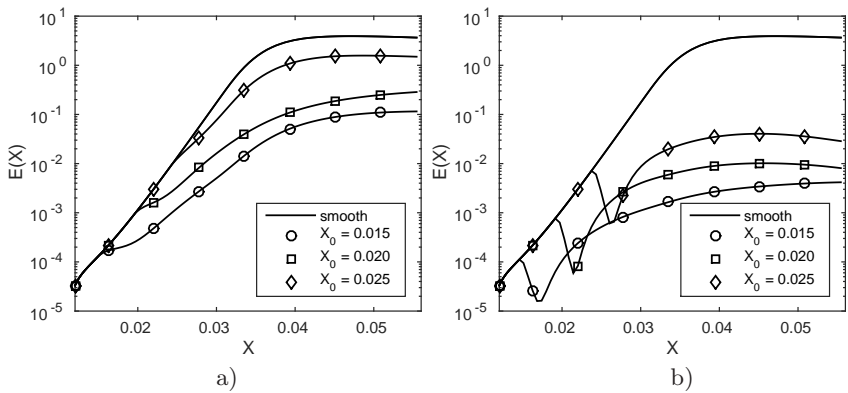

 \begin{center}
     \includegraphics[width=5.5cm]{energy_blo_lam12}
      \includegraphics[width=5.5cm]{energy_wall_lam12}\\
\hspace{7mm} a)  \hspace{55mm}  b) \\
 \end{center}
  \caption{ Energy as a function of the stream-wise direction for $\Lambda^* = 1.2$: a) control based on wall transpiration; b) control based on wall deformations.}
  \label{f9}
\end{figure}

\begin{figure}
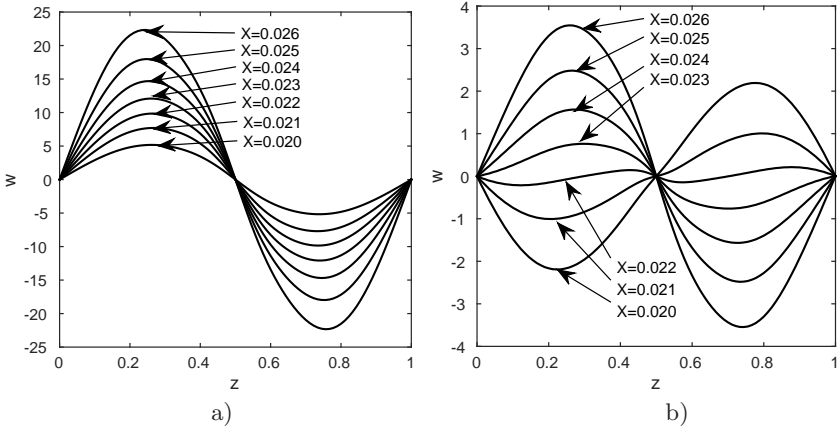

 \begin{center}
     \includegraphics[width=5.5cm]{w_X_tran}
      \includegraphics[width=5.5cm]{w_X_wall}\\
\hspace{7mm} a)  \hspace{55mm}  b) \\
 \end{center}
  \caption{ span-wise velocity component, $w$, as a function of $z$ for several stream-wise locations: a) control based on wall transpiration; b) control based on wall deformation.}
  \label{f10}
\end{figure}

In figure \ref{f9}, we compare the kinetic energy associated with the G\"{o}rtler vortex for the three control cases to the energy calculated in the uncontrolled case. The energy is significantly reduced by the control based on wall deformation; for the smallest $X_0$ location, approximately three orders of magnitude reduction is observed. The energy reduction in the case of wall transpiration is not as high as in the previous case, but it is considerable at least for the largest $X_0$ (almost two orders of magnitude). In figure \ref{f9}b, sharp minima in all energy curves associated with the controlled cases exist. These minima correspond to the stream-wise location where the low- and high-speed regions switch span-wise locations, and also to the initiation point for the transient energy growth. To further rationalize this behavior, in figure \ref{f10} we plot distributions of the span-wise velocity component, $w$, as a function of $z$ for several stream-wise locations (some upstream and some downstream of the energy minima location) at $Y=0.07$ from the wall (at the point where $w$ has a maximum in the wall-normal direction). These plots correspond to the energy curve with square symbols in figure \ref{f9} ($X_0=0.020$), for which the energy minima occurs at approximately $X=0.022$. The left hand figure corresponds to the control based on wall transpiration. We see here that both the positive and negative regions of $w$ are monotonically increasing (in the absolute value). For the control based on wall deformation on the right hand side on the other hand, there is a sign switch at both positive and negative regions. Since the curve corresponding to $X=0.022$ is the closest to the $w=0$ line, plots in figure \ref{f10} indicate that the stream-wise location where the energy minima occurs in figure \ref{f9}?corresponds to vortices that experience region switching.

\begin{figure}
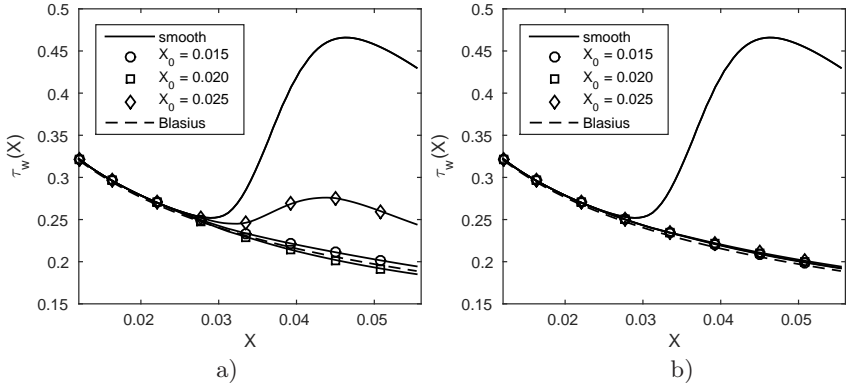

 \begin{center}
     \includegraphics[width=5.5cm]{shear_blo_lam12}
      \includegraphics[width=5.5cm]{shear_wall_lam12}\\
\hspace{7mm} a)  \hspace{55mm}  b) \\
 \end{center}
  \caption{ Span-wise averaged wall shear stress as a function of the stream-wise direction for $\Lambda^* = 1.2$: a) control based on wall transpiration; b) control based on wall deformations.}
  \label{f11}
\end{figure}

As a means to quantify the effect of control algorithms on the frictional drag, in figure \ref{f11} we plot the span-wise averaged wall shear stress for all cases, including the curve corresponding to the Blasius solution (i.e., no G\"{o}rtler vortices, and no control). First, it appears that there is a significant increase in the shear stress from approximately $X=0.032$, for G\"{o}rtler vortices in the uncontrolled case, which translates into an increase in the skin friction drag. It is obvious that the shear stress is significantly reduced by the application of either control schemes, with wall deformations being more effective; this confirms our previous results in figures \ref{f4}, \ref{f5} and \ref{f8}. The shear stress curves corresponding to the control based on wall deformations (figure \ref{f11}b) are all remarkably very close to the Blasius curve.

\begin{figure}
 \begin{center}
     \includegraphics[width=6cm]{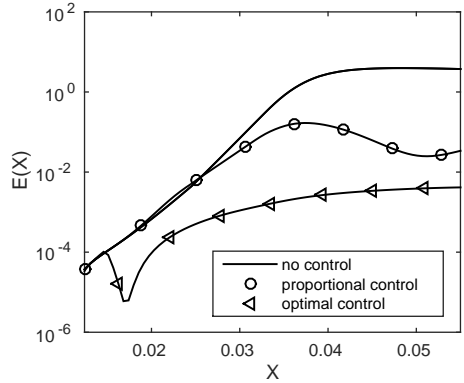}\\
 \end{center}
  \caption{ Comparison of energy plot from \cite{Sescu2} with the present result, for the same span-wise separation $\Lambda^* = 1.2$.}
  \label{f12}
\end{figure}

It is interesting to compare the energy plot obtained from the proportional control scheme that was applied in \cite{Sescu2} with the present energy result, for the same span-wise separation of $\Lambda^* = 1.2$. This is accomplished on figure \ref{f12}, which shows that the energy reduction corresponding to the optimal control is much more significant than the energy reduction corresponding to the proportional controller. This is not at all surprising, however, since the latter is based on a formulation in which the energy reduction would always be sub-optimal.

\subsection{Sensitivity to the roughness span-wise separation and diameter}

\begin{figure}
 \begin{center}
      \includegraphics[width=4.4cm]{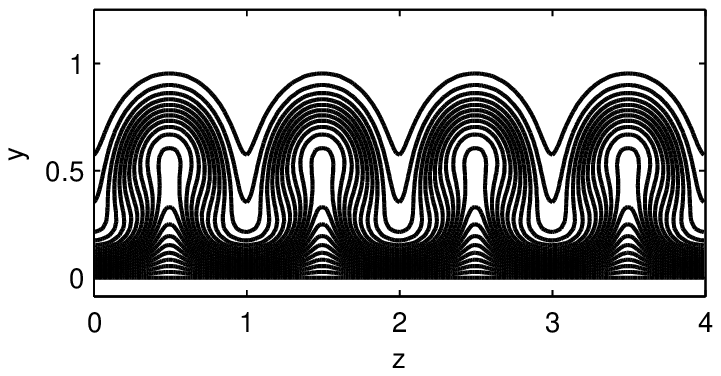} 
      \includegraphics[width=4.4cm]{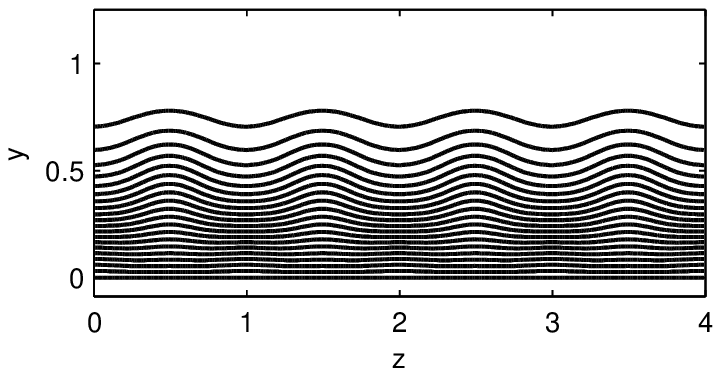} 
      \includegraphics[width=4.4cm]{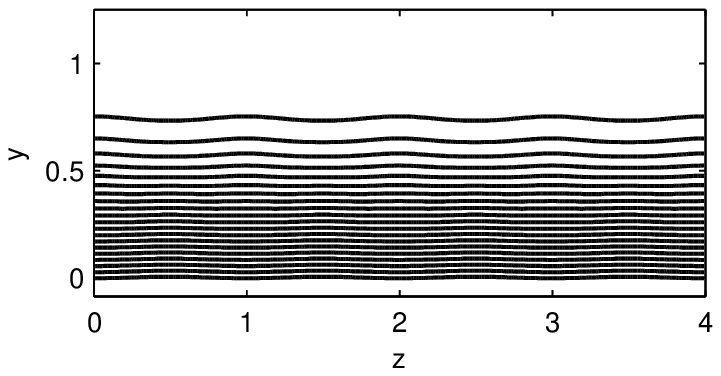}\\
\hspace{5mm}  a)  \hspace{40mm}  b)  \hspace{40mm}  c) \\
 \end{center}
  \caption{ Stream-wise velocity contours for $X_0 = 0.015$ and $\Lambda^* = 0.8$ cm: a) no control; b) control based on wall transpiration; c) control based on wall deformation.}
  \label{f13}
\end{figure}

\begin{figure}
 \begin{center}
      \includegraphics[width=4.4cm]{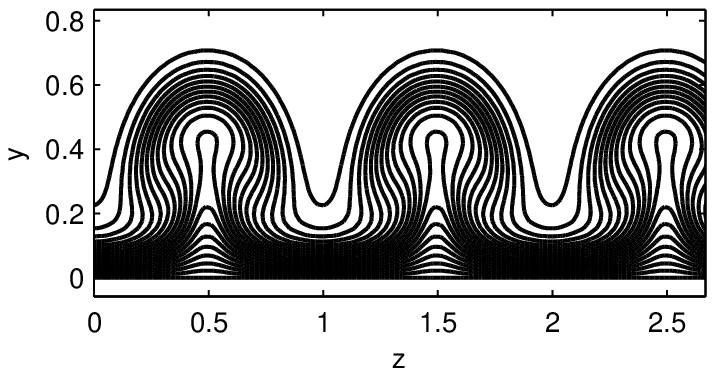} 
      \includegraphics[width=4.4cm]{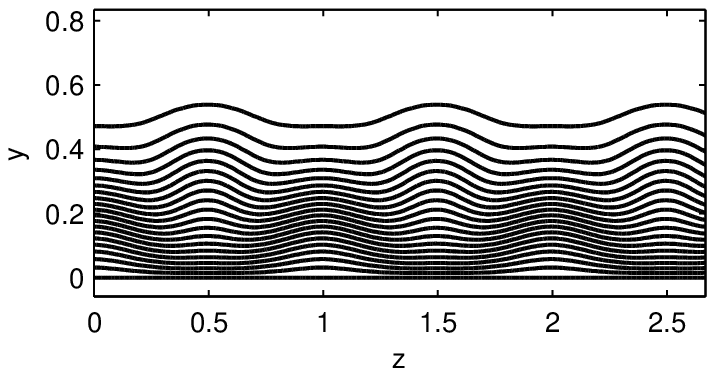} 
      \includegraphics[width=4.4cm]{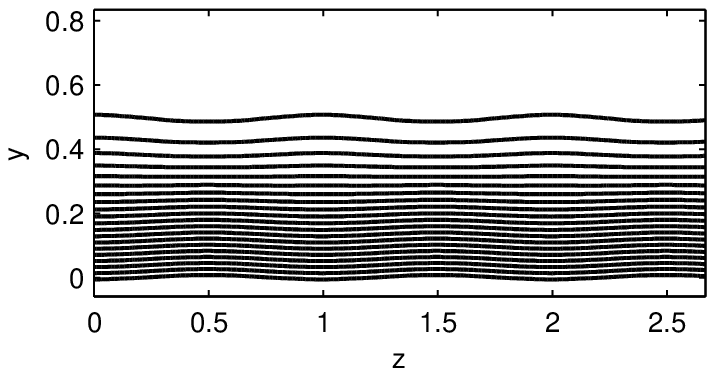}\\
\hspace{5mm}  a)  \hspace{40mm}  b)  \hspace{40mm}  c) \\
 \end{center}
  \caption{ Stream-wise velocity contours for $X_0 = 0.015$ and $\Lambda^* = 1.2$ cm: a) no control; b) control based on wall transpiration; c) control based on wall deformation.}
  \label{f14}
\end{figure}

\begin{figure}
 \begin{center}
      \includegraphics[width=4.4cm]{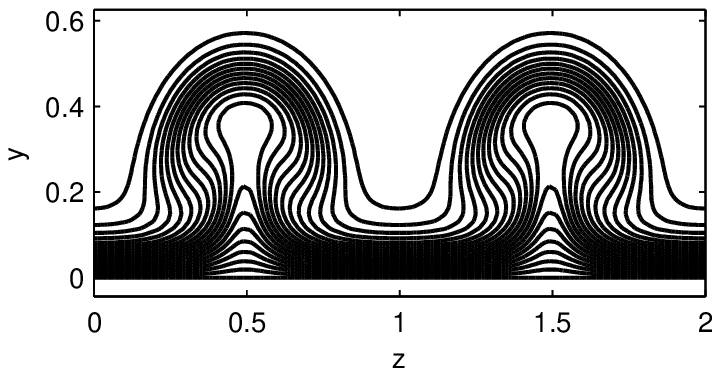} 
      \includegraphics[width=4.4cm]{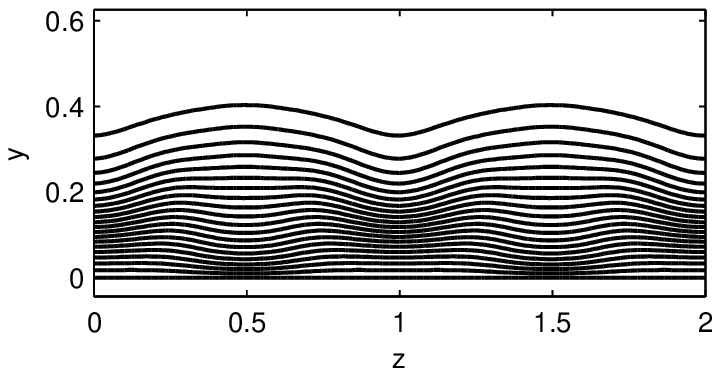} 
      \includegraphics[width=4.4cm]{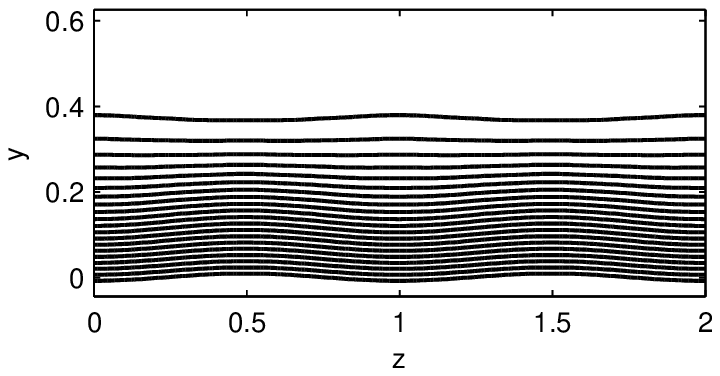}\\
\hspace{5mm}  a)  \hspace{40mm}  b)  \hspace{40mm}  c) \\
 \end{center}
  \caption{ Stream-wise velocity contours for $X_0 = 0.015$ and $\Lambda^* = 1.6$ cm: a) no control; b) control based on wall transpiration; c) control based on wall deformation.}
  \label{f15}
\end{figure}

In this section, results from three different span-wise separations, $\Lambda^*$ = 0.8, 1.2, and 1.6 cm (where the roughness diameter is fixed at $D=0.6$), and three different roughness diameters, $D=$ 0.4, 0.7, and 1.0 (where the roughness span-wise separation is fixed at $\Lambda^*$ = 1.2), are compared to each other. Based on the results from the previous subsection, the control initiation location $X_0=0.015$ has been chosen since this provided the best results in terms of both energy and shear stress distributions. Stream-wise velocity component contours in the cross plane at $X=0.04$ are shown in figures \ref{f13}, \ref{f14} and \ref{f15} for the uncontrolled and controlled vortices. Different levels of reductions in the streak amplitude are revealed by these contour plots: for the control based on wall deformations the variations in the span-wise direction are almost absent. The effect of varying the roughness diameters was less noticeable in the contour plots of the stream-wise velocity, so they are not shown here.

\begin{figure}
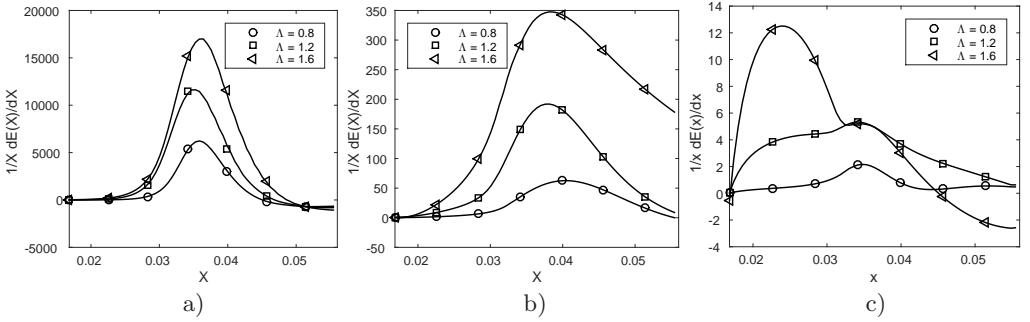
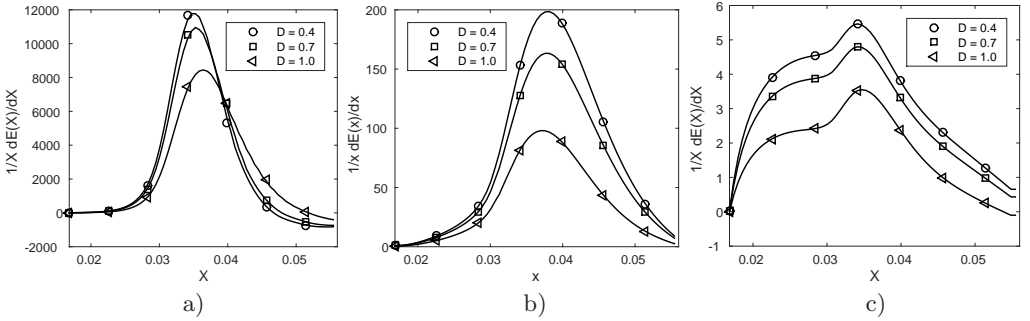

 \begin{center}
     \includegraphics[width=4.4cm]{energy_growth_smooth_lam}
     \includegraphics[width=4.4cm]{energy_growth_blo_lam}
      \includegraphics[width=4.4cm]{energy_growth_wall_lam}\\
\hspace{5mm}  a)  \hspace{40mm}  b)  \hspace{40mm}  c) \\
 \end{center}
  \caption{ Growth rate of energy as a function of the stream-wise direction: a) no control; b) control based on wall transpiration; c) control based on wall deformation.}
  \label{f16}
\end{figure}

To better highlight the difference among various control cases, the growth rate of energy, $1/X dE/dX$, is plotted here, as opposed to the energy itself as was done in the previous subsection. In figure \ref{f16} we plot the energy growth rate for different span-wise separations. One noticeable thing inferred from all cases, including the uncontrolled one, is that as the span-wise separation is increased, the energy growth rate is increased, which is all together expected because the wakes are less likely to coalesce for the largest span-wise separation. The smallest growth rates correspond to the control based on wall deformations as expected (see figure \ref{f16}c), being reduced to almost zero for the smallest span-wise separation (circles in figure \ref{f16}c). Small energy growth rate is also seen in the case of control based on wall transpiration for the smallest span-wise separation case (circles in figure \ref{f16}b). The growth rate corresponding to the largest span-wise separation in this case is an order of magnitude larger (triangles compared to circles in figure \ref{f16}b).

\begin{figure}
 \begin{center}
     \includegraphics[width=4.4cm]{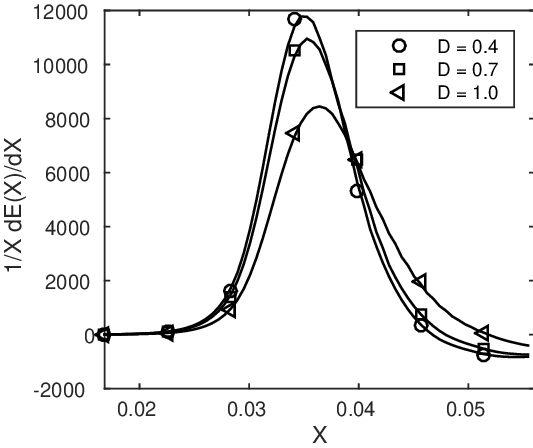}
     \includegraphics[width=4.4cm]{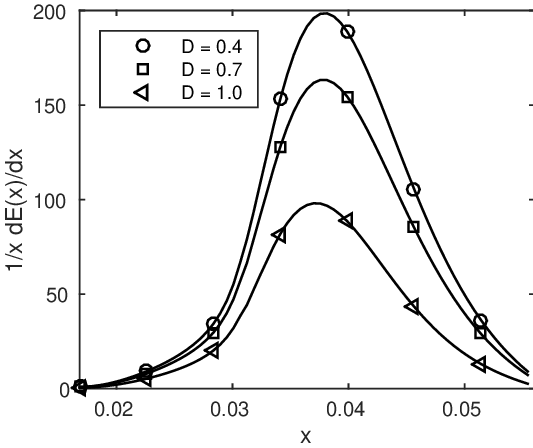}
      \includegraphics[width=4.4cm]{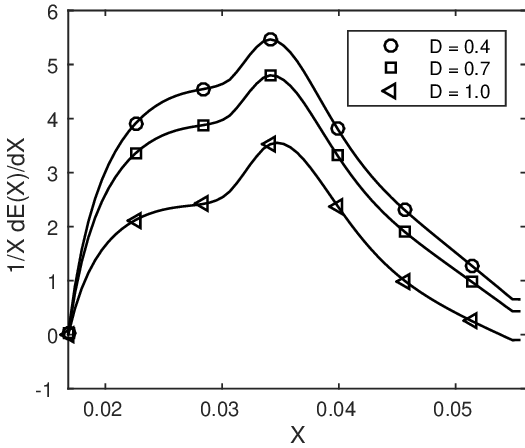}\\
\hspace{5mm}  a)  \hspace{40mm}  b)  \hspace{40mm}  c) \\
 \end{center}
  \caption{ Growth rate of energy as a function of the stream-wise direction: a) no control; b) control based on wall transpiration; c) control based on wall deformation.}
  \label{f17}
\end{figure}

In regard to the variation of the roughness diameter, figure \ref{f17} shows that the differences are not as large as the differences noticed in the previous case. As the diameter is increased the growth rates decrease for all cases, which is expected and commensurate with the results corresponding to the variation of span-wise separation (i.e., as the diameter is increased the gap between the roughness elements decreases, as long as the span-wise separation is kept constant). The growth rates corresponding to the control based on wall deformations are almost two orders of magnitude smaller than the growth rates corresponding to the control based on wall transpiration here.

\section{Discussion of the physical mechanism behind the control strategies} \label{s5}

\begin{figure}
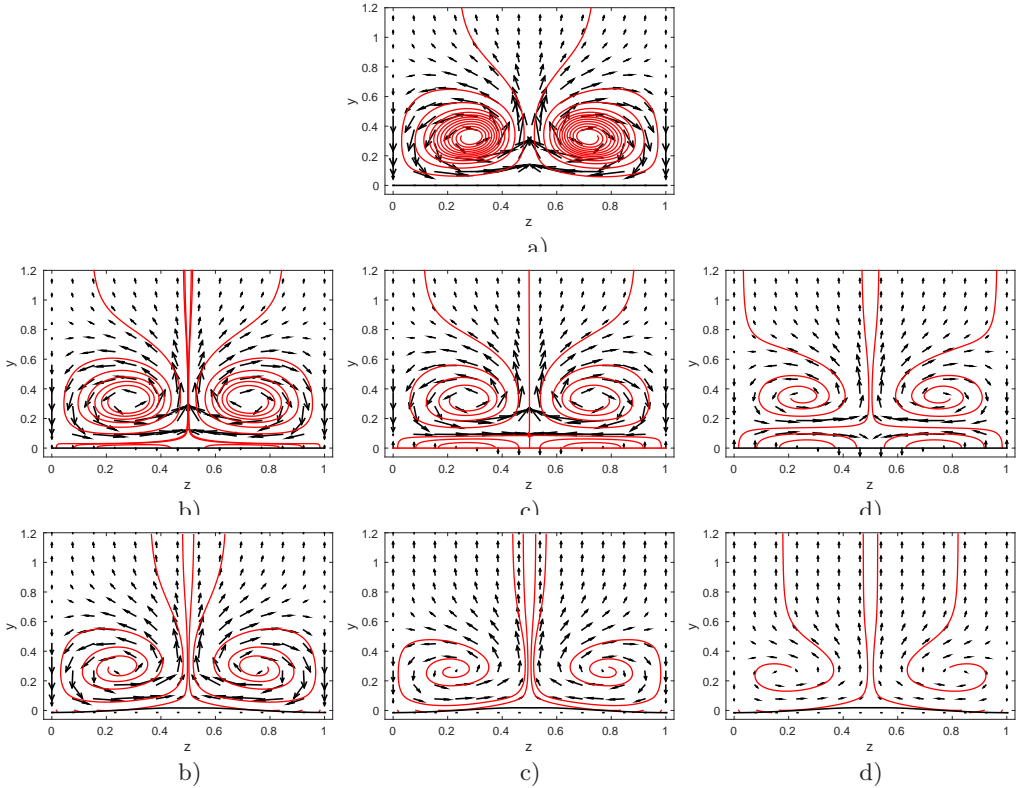

 \begin{center}
     \includegraphics[width=4.4cm]{stream_smooth_12}\\
\hspace{7mm} a)  \\
      \includegraphics[width=4.4cm]{stream1_blow_12_1}
      \includegraphics[width=4.4cm]{stream2_blow_12_1}
      \includegraphics[width=4.4cm]{stream3_blow_12_1}\\
\hspace{5mm}  b)  \hspace{40mm}  c)  \hspace{40mm}  d) \\
      \includegraphics[width=4.4cm]{stream1_wall_12_1}
      \includegraphics[width=4.4cm]{stream2_wall_12_1}
      \includegraphics[width=4.4cm]{stream3_wall_12_1}\\
\hspace{5mm}  b)  \hspace{40mm}  c)  \hspace{40mm}  d) \\
 \end{center}
  \caption{Two-dimensional vector fields $(v,w)$ and streamlines in a cross-plane $(y,z)$ located in $X=0.04$.: a) uncontrolled flow; b) wall transpiration control, iteration 1; c) wall transpiration control, iteration 5; d) wall transpiration control, iteration 10; e) wall deformation control, iteration 1; f) wall deformation control, iteration 5; g) wall deformation control, iteration 10.}
  \label{f18}
\end{figure}

In this section, we discuss the main physical mechanisms that are behind the control based on wall transpiration and wall deformation, and explain why the latter is more effective than the former in reducing the vortex energy. In figure \ref{f18}, we plot two-dimensional vector fields $(v,w)$ and streamlines in a cross-plane $(y,z)$ located in $X=0.04$, for the span-wise separation $\Lambda^* = 1.2$. Two streamlines were seeded in close proximity to the centers of two vortices, respectively, while other streamlines were seeded from the wall. These latter ones have meaning for control based on wall transpiration because the velocity is non-zero at the wall as a result of blowing and suction. The top plot (figure \ref{f18}a) represents the uncontrolled flow, and it shows two counter-rotating vortices lifting low momentum from the wall at the span-wise location between the two vortices, as well as distributing high momentum fluid from the upper layers to the wall at the $z=0$ or $z=1$ planes. It is well-known that this exchange of high- and low-momentum in the vertical direction brings about variations of the stream-wise velocity along the span-wise direction, which is the main driver of high- and low-speed streaks. 

Figures \ref{f18}b and \ref{f18}c represent the vortices for the control based on wall deformation and transpiration, respectively, after the first iteration in the control algorithm. The other parts of the figure (\ref{f18}d, \ref{f18}e,\ref{f18}f and \ref{f18}g) represent the vortices at other two subsequent iterations as indicated in the figure caption. As the number of iterations in the control algorithm increases, a weakening of the vortex strength is apparent as a result of both control algorithms. The strength of the vortex is qualitatively measured here by the length of the arrows of the two-dimensional vector field $(v,w)$. The 'relaxation' of vortices is also revealed by the evolution of the streamline that is seeded from a location that is close to the center of the vortex. For the uncontrolled flow in figure \ref{f18}a, it is noted that these streamlines are very close to each other; as the iteration number increases, however, the streamlines become more scarce and distant from one another (this is more prevalent for the control based on wall deformation). For the control based on wall transpiration in figures \ref{f18}b, \ref{f18}d and \ref{f18}f, an additional weak vortical flow is initiated in the vicinity of the wall due to a span-wise alternation of blowing and suction. This forces the two counter-rotating vortices to move upward, changing the vertical location of the G\"{o}rtler vortex.

\begin{figure}
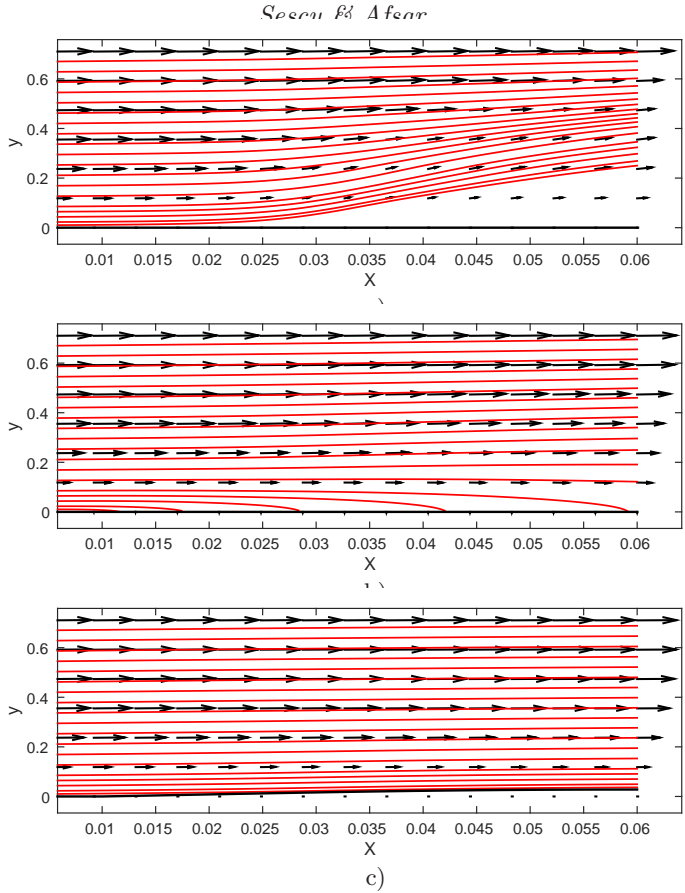

 \begin{center}
     \includegraphics[width=9cm]{stream_xy2_smooth_12} \\
\hspace{8mm} a)  \\
     \includegraphics[width=9cm]{stream_xy2_blow_12_1} \\
\hspace{8mm} b)  \\
     \includegraphics[width=9cm]{stream_xy2_wall_12_1}  \\
\hspace{8mm} c)  \\
 \end{center}
  \caption{Two-dimensional vector fields $(u,v)$ and streamlines in a sectional plane $(X,y)$ located in $z = 0.5$ (where low momentum is lifted from the wall): a) uncontrolled flow; b) wall transpiration control; c) wall deformation control.}
  \label{f19}
\end{figure}

In figures \ref{f19} and \ref{f20}, we plot two-dimensional vector fields and streamlines for uncontrolled and controlled boundary layers, in sectional planes $(x,y)$ that are located at $z=0.5$ and $z=0$, respectively. In figure \ref{f19}a, corresponding to the uncontrolled flow, streamlines that are seeded in close proximity to the wall reveal how fluid particles carrying low momentum are lifted up from the lower layers to the upper layers. The deceleration of fluid particles can be noted by inspecting the variation of the length of arrows that form the vector field (the length of the arrows is proportional to the velocity magnitude in its location): by following the horizontal line of vectors starting from the left boundary at approximately $y=0.36$ in figure \ref{f19}a, one can notice how the length of arrows decreases as the fluid particle advances downstream. Now, if one inspects the same line of vectors in figure \ref{f19}b or \ref{f19}c (initiated at $y=0.36$ from the left), it is seen that the length of the arrow does not change much as the fluid particle moves downstream (there is a very small decrease for control based on wall transpiration in figure \ref{f19}b). The streamline at this wall-normal location ($y=0.36$) is also parallel to the wall for both control cases, while the same streamline in the uncontrolled case (figure \ref{f19}a) shows a significant deviation in the upward direction. The same conclusions can be reached for the line of vectors located at $y=0.12$ (the streamline associated with this location is parallel to the wall as well). Below this level ($y=0.12$), however, there are differences between the two control algorithms: first, the suction effect is revealed in figure \ref{f19}b by the first five streamlines changing direction and ending at the wall because the fluid particles are absorbed into the wall; second, figure \ref{f19}c shows a slight deviation of the wall line in the upward direction as a results of the control scheme, which accelerates fluid particle, thus increasing the momentum in the vicinity of the wall (all streamlines in this case become parallel to the wall).

In figure \ref{f20}, the same picture as in figure \ref{f19} is shown, except that another span-wise location corresponding to the downwelling of fluid particle from upper layers to the wall is considered. As mentioned before, at this span-wise location $z=0$ high momentum from above is brought downward, which can be clearly noticed by the increase in the length of arrows along the line of vectors that starts in approximately $y=0.12$ or $y=0.25$ from the left boundary. Shown in figure \ref{f20}a are also streamlines that are deviated towards the wall as a result of the counter-rotation posed by the vortices. The effect of blowing at this span-wise location is seen in figure \ref{f20}b, where the streamlines are deviated from the wall; this brings about changes in the upper layers, where fluid particles are constrained to move at a constant speed and in a trajectory that is parallel to the wall (see the line of vectors that starts in $y=0.25$, for example). In figure \ref{f20}c, the downward deviation of the wall geometry can be seen, which slows down the fluid particles gently, thus decreasing the momentum in the proximity to the wall; again, this yields a constant velocity along the vector lines, with streamlines that remain parallel to the wall.

\begin{figure}
 \begin{center}
     \includegraphics[width=9cm]{stream_xy1_smooth_12} \\
\hspace{8mm} a)  \\
     \includegraphics[width=9cm]{stream_xy1_blow_12_1} \\
\hspace{8mm} b)  \\
     \includegraphics[width=9cm]{stream_xy1_wall_12_1} \\
\hspace{8mm} c)  \\
 \end{center}
  \caption{Two-dimensional vector fields $(u,v)$ and streamlines in a sectional plane $(X,y)$ located in $z = 0$ (where low momentum is brought to the wall): a) uncontrolled flow; b) wall transpiration control; c) wall deformation control.}
  \label{f20}
\end{figure}

\begin{figure}
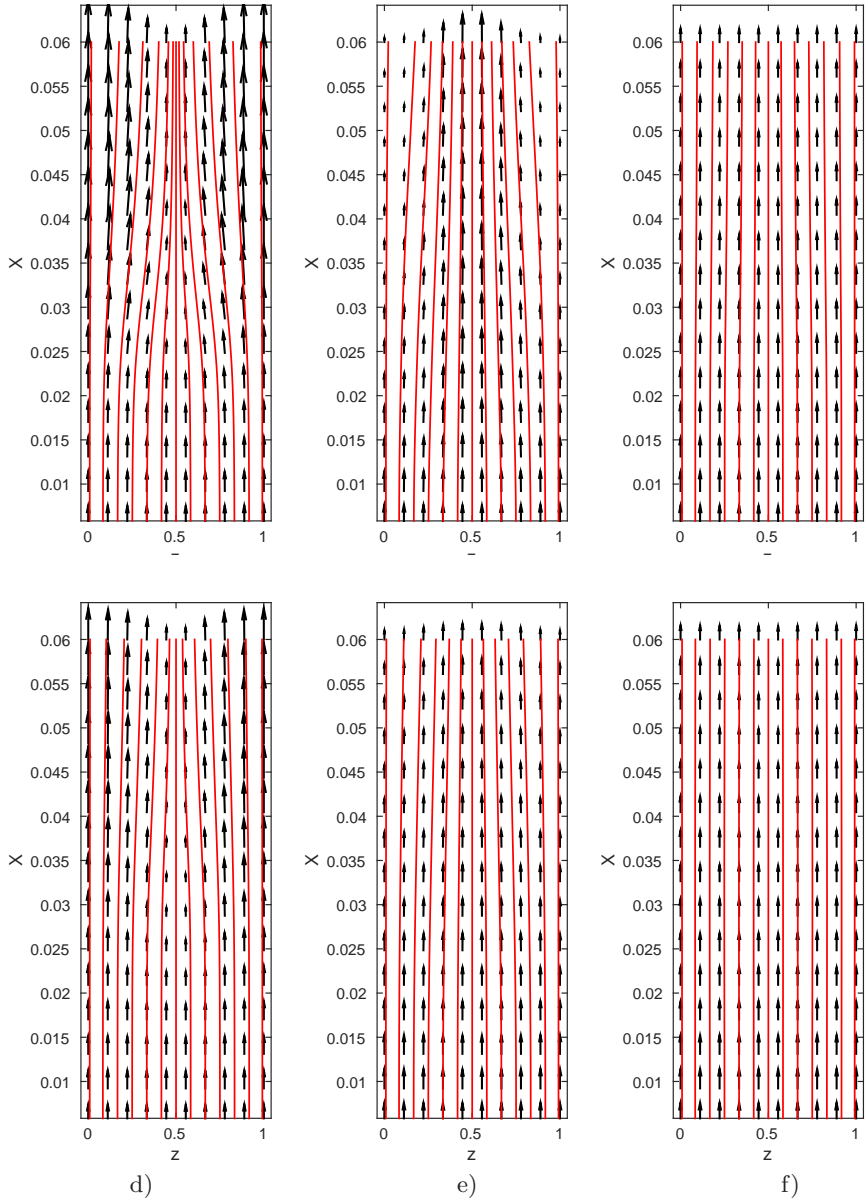

 \begin{center}
     \includegraphics[width=3.5cm]{stream_xz1_smooth_12} \hspace{2mm}
      \includegraphics[width=3.5cm]{stream_xz1_blow_12_1} \hspace{2mm}
      \includegraphics[width=3.5cm]{stream_xz1_wall_12_1}\\
\hspace{7mm} a)  \hspace{38mm}  b)   \hspace{38mm}  c) \\
     \includegraphics[width=3.5cm]{stream_xz2_smooth_12} \hspace{2mm}
      \includegraphics[width=3.5cm]{stream_xz2_blow_12_1} \hspace{2mm}
      \includegraphics[width=3.5cm]{stream_xz2_wall_12_1}\\
\hspace{7mm} d)  \hspace{38mm}  e)   \hspace{38mm}  f) \\
 \end{center}
  \caption{Two-dimensional vector fields $(u,w)$ and streamlines in a sectional plane $(X,z)$ parallel to the wall: a) uncontrolled flow, $y = 0.02$; b) wall transpiration control, $y = 0.02$; c) wall deformation control, $y = 0.02$; d) uncontrolled flow, $y = 0.15$; e) wall transpiration control, $y = 0.15$; d) wall deformation control, $y = 0.15$.}
  \label{f21}
\end{figure}

Finally, in figure \ref{f21} we plot two-dimensional vector fields $(u,w)$ and streamlines in two sectional planes $(x,z)$ that are parallel to the wall, one very close to the wall ($y=0.02$) and the other at $y=0.15$ (the reason for choosing the latter becomes clear when looking at vectors and streamline plots in figures \ref{f19} and \ref{f20}; at this elevation the velocity is constant along a horizontal line, and the streamline is parallel to the wall). The uncontrolled flow in figure \ref{f21}a, corresponding to the plane $y=0.02$, shows a deviation of fluid particles towards $z=0.5$ line, which is the effect of counter-rotating vortices; this produces a small cross-flow component. As a result of the control, as shown in figures \ref{f21}b and \ref{f21}c, the fluid particles are brought back to their original upstream trajectories, with the cross flow component being negligible (note that a small cross-flow component is still seen in the plot corresponding to the control based on wall transpiration). At the other elevation ($y=0.15$), there is somewhat smaller cross-flow component in the uncontrolled case (figure \ref{f21}d) as expected, and no cross-flow component for both control cases.

Different mechanisms towards controlling a boundary layer flow distorted by centrifugal instabilities are at play here for the two schemes introduced in this study:
\begin{itemize}

\item for control based on wall deformation, a positive displacement of the wall (at the span-wise location between the two vortices, as seen figure \ref{f19}c) accelerates fluid particles in the stream-wise direction, while negative displacement (see figure \ref{f20}c) decelerates them. This means that the momentum of fluid particles in the close proximity to the wall (in between the vortices) is increased, thus weakening the effect of lift-up of low momentum fluid particles from the wall, as previously described. The opposite occurs at negative displacement locations, where the momentum of fluid particles at the wall is decreased. There is also an effect (although small) on the cross-flow velocity component as shown in figure \ref{f21}. 

\item the mechanism behind the control based on wall transpiration closely resembles the opposition control type, as introduced and described in previous studies concerning turbulent boundary layers or channel flows (see, for example, \cite{Choi}, \cite{Koumoutsakos1,Koumoutsakos2}, \cite{Stroh}), although, here, the sensor is not the vertical velocity measured in a plane parallel to the wall. However, as seen in figures \ref{f18}b, \ref{f18}d and \ref{f18}f, blowing is applied in regions where the vertical velocity in the flow is negative, and suction is applied where the vertical velocity in the flow is positive; this is the main mechanism of reducing the energy of the G\"{o}rtler vortices. Compared to the control based on wall deformation, the effect on the cross-flow velocity is more significant as shown in figure \ref{f18}f, where the vortical flow induced by alternating blowing and suction contributes to the increase of the cross-flow.

\end{itemize}

We emphasize again that there are significant differences between steady G\"{o}rtler vortices considered here and wall turbulence considered in previous studies that might explain why for wall turbulence it was found that the control based on wall transpiration can be more effective than wall deformation. As mentioned previously, we solve a parabolic problem in steady state, where we assume variations in the stream-wise direction to be small for both the flow and wall geometry. One of the reasons that the wall deformation control is more effective than the wall transpiration control, in our study, is clear from the previous discussions of vector fields and streamlines (figures \ref{f18}-\ref{f21}): in a nutshell, the alternation of blowing and suction in the span-wise direction for the former type of control introduce an additional vortical flow that, on one hand, deviates or attract streamlines from/to the wall, and, on the other hand, influences the cross-flow in proximity to the wall; this is not the case for the control based on wall deformation, where it was found that the mechanism of flow alteration is different. Moreover, in the control based on wall transpiration case, the additional vortical flow induced in the vicinity of the wall forces G\"{o}rtler vortices to move slightly upward, thus preventing their complete elimination.

\section{Summary and conclusions}\label{s6}

An optimal control strategy in the high Reynolds number asymptotic framework was derived and tested on G\"{o}rtler vortices developing along a concave surface. The Navier-Stokes equations were simplified to obtained the boundary region equations (BRE), which is a parabolic set of equations since the stream-wise pressure derivative and the second derivatives were neglected. The effect of wall deformations was incorporated into the model by a Prandtl transformation applied to the BRE, which were then solved numerically using a marching algorithm. The variations in local surface shape then allowed for a relatively straightforward control strategy to be implemented to the transformed set of equations in order to determine the optimum wall deformation or transpiration that reduced the energy of the vortices. G\"{o}rtler vortices were initiated by perturbing the upstream flow with a periodic array of roughness elements placed near the leading edge, using a previously derived asymptotic solution to obtain the upstream boundary conditions (\cite{Goldstein1,Goldstein2}). The vortex energy was controlled via an optimal control algorithm based on Lagrange multipliers and adjoint equations, where the wall displacement or the wall transpiration velocity serve as control variables. The modified BREs eqns (\ref{neq1})-(\ref{neq4}) explicitly included an arbitrary function $\mathscr{F}(X,z)$ representing the local surface deformation, and therefore provided a parametric mechanism for the optimal control. The optimal control strategy was aimed at minimizing the difference between the calculated wall shear stress and the Blasius shear stress.  The cost functional was defined in terms of the energy associated with the G\"{o}rtler vortices, and the sensors were functions of the wall shear stress. Local wall deformations appeared to resemble elongated surface shapes, in the form of streaks, that we optimized to control the G\"{o}rtler vortex energy. 

An extensive numerical study of the effect of controlled wall deformation versus transpiration on the vortex development for incompressible boundary layer flow over a curved wall has been performed. Various results in terms of contour plots of stream-wise velocity, energy distribution in the stream-wise direction, or distributions of the span-wise average shear stress showed that the optimal control algorithm is very effective in counteracting the G\"{o}rtler vortices. The kinetic energy associated with the vortices was shown to decrease by almost three orders of magnitude from the original amplitude in some cases, with a commensurate reduction in the wall shear stress. Our results clearly indicate that the controlled surface deformations are more effective than wall transpiration in altering the stream-wise development of the G\"{o}rtler vortices for a given roughness height and span-wise wavelength. Potential explanations for this difference were given, and the physical mechanism behind the control was described. While the boundary layer was excited by a row of roughness elements as derived in \cite{Goldstein1}, the algorithm can be easily extended to include freestream disturbances or turbulence at the upstream boundary.

The  proposed control algorithm has potential application to a number of realistic problems including aircraft wing design, to mitigate induced drag, or in supersonic wind tunnels, where it is known that flow instabilities in the test section are triggered by G\"{o}rtler vortices developing on the concave portion of the walls. Of course, the latter application will necessitate the extension of the algorithm in the compressible regime, which is a matter of future analysis.


\section*{Acknowledgments}

A.S. would like to acknowledge support from NASA EPSCoR RID Program through Mississippi State Grant Consortium directed by Dr. Nathan Murray. MZA would like to thank Strathclyde University for financial support from the Chancellor's Fellowship.


\section*{APPENDIX A: Adjoint equations}

In the particular case considered in this study (with $\psi = \mathscr{F}$ representing the wall deformation, and $\phi = v_w$ representing the wall transpiration), the Lagrangian can be expanded as

\begin{eqnarray}\label{opt}
&& \mathscr{L}(u,v,w,p,\mathscr{F},v_w,u^a,v^a,w^a,p^a,s)   \nonumber  \\
&=& \frac{\sigma_1}{2} \int_{X_0}^{X_1} \int_{\Omega} \|\frac{\partial \mathscr{F}}{\partial X}\|^2 + \|\mathscr{F}\|^2 d\Omega dX 
+   \frac{\sigma_2}{2} \int_{X_0}^{X_1} \int_{\Gamma} \|\frac{\partial v_w}{\partial X}\|^2 + \|v_w\|^2 d\Gamma dX  \nonumber \\
&-&  \int_{X_0}^{X_1} \int_{\Gamma} s \left( v - v_w \right) d\Gamma dX
- \int_{0}^{X_0} \int_{\Gamma} s v d\Gamma dX
- \int_{X_1}^{X_t} \int_{\Gamma} s v  d\Gamma dX  + \mathscr{H}
\end{eqnarray}
where

\begin{eqnarray}\label{}
 \mathscr{H}
&=&  \frac{\alpha}{2} \int_{X_{s0}}^{X_{s1}} \int_{\Gamma} \|\tau_w - \tau_0\|^2 d\Gamma dX  \nonumber \\
&-&  \int_{0}^{X_t} \int_{\Omega} u^a \left[ u \frac{\partial u}{\partial X}+\hat{v}\frac{\partial u}{\partial Y}+w\frac{\partial u}{\partial z} - \frac{\partial^2 u}{\partial Y^2} - \mathscr{D}^2 u + \frac{\partial \mathscr{F}^2}{\partial z^2} \frac{\partial u}{\partial Y} \right] d\Omega dX   \nonumber \\
&-&  \int_{0}^{X_t} \int_{\Omega} v^a \left[ u \frac{\partial v}{\partial X}+\hat{v}\frac{\partial v}{\partial Y}+w\frac{\partial v}{\partial z}+G_{\Lambda}u^{2} + \frac{\partial p}{\partial Y} - \frac{\partial^2 v}{\partial Y^2} - \mathscr{D}^2 v + \frac{\partial \mathscr{F}^2}{\partial z^2} \frac{\partial v}{\partial Y} \right] d\Omega dX  \nonumber \\
&-&  \int_{0}^{X_t} \int_{\Omega} w^a \left[ u \frac{\partial w}{\partial X}+\hat{v}\frac{\partial w}{\partial Y}+w\frac{\partial w}{\partial z} + \mathscr{D} p - \frac{\partial^2 w}{\partial Y^2} - \mathscr{D}^2 w + \frac{\partial \mathscr{F}^2}{\partial z^2} \frac{\partial w}{\partial Y} \right] d\Omega dX  \nonumber \\
&-&  \int_{0}^{X_t} \int_{\Omega} p^a \left[ \frac{\partial u}{\partial X}+\frac{\partial \hat{v}}{\partial Y}+\frac{\partial w}{\partial z} \right] d\Omega dX.  \nonumber 
\end{eqnarray}
which is a notation that is introduced to avoid repetitions of this quantity in the subsequent equations. In the above equations, the control is applied in specified intervals $[X_0,X_1]$ only, $\Omega$ is the cross-section domain $[0,\infty] \times [z_1,z_2]$ ranging from the wall ($Y=0$) to infinity and from $z_1$ to $z_2$ in the span-wise direction, $\Gamma$ is the wall boundary line for a given $X$, $\tau_w$ is the wall shear stress, $\tau_0$ is a target shear stress (equal to the value corresponding to the Blasius solution), and $[X_{s1},X_{s2}]$ is the interval where the cost function is defined. The last three integrals in (\ref{opt}) are used to enforce the boundary condition on $v$, which includes the wall transpiration. If we take the directional derivative of the Lagrangian with respect to $p$, the result is

\begin{equation}\label{}
\int_{0}^{X_{t}} \int_{\Omega} \left[ v^a \frac{\partial (\delta p)}{\partial Y} + w^a \mathscr{D} (\delta p)  \right] d\Omega dX = 0
\end{equation}
Then, by applying an integration by parts in $\Omega$ we obtain 

\begin{equation}\label{}
\int_{0}^{X_t} \int_{\Gamma} \left[ v^a \delta p + w^a \delta p \left(1-\frac{\partial \mathscr{F}}{\partial z} \right)  \right]  d\Gamma dX - \int_{0}^{X_{t}} \int_{\Omega} \delta p \left( \frac{\partial v^a}{\partial Y} + \mathscr{D} w^a  \right) d\Omega dX = 0
\end{equation}
and assuming arbitrary variations of $\delta p$, the first adjoint equation is obtained in the form

\begin{equation}\label{ad1}
\frac{\partial v^a}{\partial Y} + \mathscr{D} w^a = 0  \hspace{3mm} on  \hspace{3mm} [0,X_t] \times \Omega,
\end{equation}
where $\delta p|_{\Gamma}=0$.

Next, we take the directional derivative of the Lagrangian with respect to $u$, and obtain

\begin{eqnarray}\label{}
 \int_{0}^{X_{t}} \int_{\Omega} && \left[ u^a \delta u \frac{\partial u}{\partial X} + u^a u \frac{\partial (\delta u)}{\partial X} + u^a v \frac{\partial (\delta u)}{\partial Y} - \frac{\partial \mathscr{F}}{\partial X} u^a \delta u \frac{\partial u}{\partial Y} - \frac{\partial \mathscr{F}}{\partial X} u^a u \frac{\partial (\delta u)}{\partial Y}  \right.  \nonumber \\ 
 &-& \left.  \frac{\partial \mathscr{F}}{\partial z} u^a w \frac{\partial (\delta u)}{\partial Y} + u^a w \frac{\partial (\delta u)}{\partial z} - u^a \frac{\partial^2 (\delta u)}{\partial Y^2} - u^a \mathscr{D}^2(\delta u) \right.  \nonumber \\
 &+& \left. \frac{\partial^2 \mathscr{F}}{\partial z^2} u^a \frac{\partial (\delta u)}{\partial Y} + v^a \delta u \frac{\partial v}{\partial X} -  \frac{\partial \mathscr{F}}{\partial X} v^a \delta u \frac{\partial v}{\partial Y} + 2 G_{\Lambda} v^a u \delta u \right.  \nonumber \\
 &+& \left. w^a \delta u \frac{\partial w}{\partial X} - \frac{\partial \mathscr{F}}{\partial X} w^a \delta u \frac{\partial w}{\partial Y} + p^a \frac{\partial \mathscr{F}}{\partial X} - \frac{\partial \mathscr{F}}{\partial X} p^a \frac{\partial (\delta u)}{\partial Y} \right] d\Omega dX = 0
\end{eqnarray}
where terms involving $\partial \mathscr{F}/\partial X$ or $\partial \mathscr{F}/\partial z$ derive from the dependence of $\hat{v}$ on $u$, that is $\hat{v} = v - \left( u \partial \mathscr{F}/\partial X + w \partial \mathscr{F}/\partial z \right)$. Integrating by parts in $\Omega$ or $[0,X_t]$, 

\begin{eqnarray}\label{}
&& \int_{\Omega}  \left[ (u^a u \delta u) |_{0}^{X_t} + (p^a \delta u) |_{0}^{X_t} \right] d\Omega  \nonumber \\
&+& \int_{0}^{X_{t}} \int_{\Gamma}  \left[ u^a v \delta u - \frac{\partial \mathscr{F}}{\partial X} u^a u \delta u - \frac{\partial \mathscr{F}}{\partial z} u^a w \delta u + u^a w \delta u - u^a \frac{\partial (\delta u)}{\partial Y} +  \frac{\partial u^a}{\partial Y} \delta u  \right.  \nonumber \\ 
 &-& \left. u^a \frac{\partial (\delta u)}{\partial z} + \frac{\partial u^a}{\partial z} \delta u - 2 \frac{\partial \mathscr{F}}{\partial z} \left[ u^a \frac{\partial (\delta u)}{\partial Y} - \frac{\partial u^a}{\partial z} \delta u \right] + \left( \frac{\partial \mathscr{F}}{\partial z} \right)^2 \left[ u^a \frac{\partial (\delta u)}{\partial Y} - \frac{\partial u^a}{\partial Y} \delta u \right] \right.  \nonumber \\
 &+& \left. \frac{\partial^2 \mathscr{F}}{\partial z^2} u^a \delta u
 - \frac{\partial \mathscr{F}}{\partial X} p^a \delta u \right] d\Gamma dX = 0  \nonumber \\
 &+&
 \int_{0}^{X_{t}} \int_{\Omega} \delta u \left[ u^a  \frac{\partial u}{\partial X} - u  \frac{\partial u^a}{\partial X} - v  \frac{\partial u^a}{\partial Y} - \frac{\partial \mathscr{F}}{\partial X} u^a  \frac{\partial u}{\partial Y} + \frac{\partial \mathscr{F}}{\partial X} u  \frac{\partial u^a}{\partial Y}  \right.  \nonumber \\ 
 &+& \left.  \frac{\partial \mathscr{F}}{\partial z} w  \frac{\partial u^a}{\partial Y} - w  \frac{\partial u^a}{\partial z} -  \frac{\partial^2 u^a}{\partial Y^2} -  \mathscr{D}^2(u^a)
 - \frac{\partial^2 \mathscr{F}}{\partial z^2}  \frac{\partial u^a}{\partial Y} + v^a  \frac{\partial v}{\partial X} -  \frac{\partial \mathscr{F}}{\partial X} v^a  \frac{\partial v}{\partial Y} + 2 G_{\Lambda} v^a u  \right.  \nonumber \\
 &+& \left. w^a  \frac{\partial w}{\partial X} - \frac{\partial \mathscr{F}}{\partial X} w^a  \frac{\partial w}{\partial Y} -  \frac{\partial p^a}{\partial X} + \frac{\partial \mathscr{F}}{\partial X}  \frac{\partial p^a}{\partial Y} \right] d\Omega dX = 0  \nonumber \\ 
\end{eqnarray}
where the first integral is obtained from integration by parts in $[0,X_t]$, and the second integral is obtained from integration by parts in $\Omega$; some additional terms have appeared from integrating by parts two times, wherever second derivatives of $\delta u$ are involved (i.e., for two generic functions $f$ and $g$, $\int f g'' dx = f g' - f' g + \int f'' g dx$). Thus, the second adjoint equation is obtained in the form

\begin{eqnarray}\label{ad2}
&-& u\frac{\partial u^a}{\partial X} - \hat{v}\frac{\partial u^a}{\partial Y} - w\frac{\partial u^a}{\partial z} + u^a \frac{\partial u}{\partial X}+v^a \frac{\partial v}{\partial X}+w^a \frac{\partial w}{\partial X} -  \frac{\partial \mathscr{F}}{\partial X} (u^a \frac{\partial u}{\partial Y}+v^a \frac{\partial v}{\partial Y}+w^a \frac{\partial w}{\partial Y})   \nonumber \\
&+& 2 G_{\Lambda}u v^a
- \frac{\partial p^a}{\partial X} + \frac{\partial \mathscr{F}}{\partial X} \frac{\partial p^a}{\partial Y} - \frac{\partial^2 u^a}{\partial Y^2} - \mathscr{D}^2 u^a - \frac{\partial^2 \mathscr{F}}{\partial z^2} \frac{\partial u^a}{\partial Y} = 0
\end{eqnarray}
where $\delta u|_{\Gamma}=0$, $\delta u|_{0}^{X_t}=0$, and its first derivatives with respect to $y$ or $z$ are zero at the boundaries. 

Similarly, the third and the fourth adjoint equations, corresponding to the y- and z-momentum equations from BRE, respectively, are obtained in the form:

\begin{eqnarray}\label{ad3}
&-& u\frac{\partial v^a}{\partial X} - \hat{v}\frac{\partial v^a}{\partial Y} - w\frac{\partial v^a}{\partial z} + u^a \frac{\partial u}{\partial Y}+v^a \frac{\partial v}{\partial Y}+w^a \frac{\partial w}{\partial Y} - \frac{\partial p^a}{\partial Y} - \frac{\partial^2 v^a}{\partial Y^2} \nonumber \\ 
&-& \mathscr{D}^2 v^a - \frac{\partial^2 \mathscr{F}}{\partial z^2} \frac{\partial v^a}{\partial Y} = 0
\end{eqnarray}
\begin{eqnarray}\label{ad4}
&-& u\frac{\partial w^a}{\partial X} - \hat{v}\frac{\partial w^a}{\partial Y} - w\frac{\partial w^a}{\partial z} + u^a \frac{\partial u}{\partial z}+v^a \frac{\partial v}{\partial z}+w^a \frac{\partial w}{\partial z} -  \frac{\partial \mathscr{F}}{\partial z} (u^a \frac{\partial u}{\partial Y}+v^a \frac{\partial v}{\partial Y}+w^a \frac{\partial w}{\partial Y})   \nonumber \\
&-& \mathscr{D} p^a - \frac{\partial^2 w^a}{\partial Y^2} - \mathscr{D}^2 w^a - \frac{\partial^2 \mathscr{F}}{\partial z^2} \frac{\partial w^a}{\partial Y} = 0
\end{eqnarray}
respectively. Upon deriving the third adjoint equation (\ref{ad3}), an expression is obtained for $s$ (the Lagrange multiplier with respect to the transpiration condition; see equation \ref{opt}) as a result of the integration by parts:

\begin{eqnarray}\label{ssa}
s = -p^a - v^a (u + \hat{v} + w) - \left[ 1 + \left( \frac{\partial \mathscr{F}}{\partial z} \right)^2 \right] v^a_Y + 2 \frac{\partial \mathscr{F}}{\partial z} v^a_Y + \frac{\partial^2 \mathscr{F}}{\partial z^2} v^a. \nonumber
\end{eqnarray}

The initial and boundary conditions associated with the adjoint equations are

\begin{equation}\label{}
(u^a,v^a,w^a,p^a)|_{X=X_t} = (0,0,0,0) \hspace{2mm} in  \hspace{2mm} \Omega,
\end{equation}
\begin{equation}\label{}
 (u^a,v^a,w^a)|_{\Gamma} = 
  \begin{cases}
  (\alpha(\tau_w - \tau_0),0,0) \hspace{2mm} for  \hspace{2mm} X \in [X_{s0},X_{s1}] &  \\
  (0,0,0) \hspace{2mm} otherwise  & 
  \end{cases}
\end{equation}
and 

\begin{equation}\label{}
(u^a,v^a,w^a,p^a)|_{Y \rightarrow \infty} = (0,0,0,0) \hspace{2mm}
\end{equation}

In the present study the optimal control is performed with either wall transpiration or wall deformation (and not with both at the same time), so one of the two first integrals is (\ref{opt}) is set to zero, depending on the type of control employed (also, the last three integrals in the Lagrangian are not necessary for control based on wall deformation). In this study, the values of the constant factors $\alpha$, $\sigma_1$ and $\sigma_2$ are $1$, $0.7$ and $0.1$, respectively.

\section{APPENDIX B: Optimality condition for control based on wall transpiration}

With wall transpiration, $v_w$, considered as the objective variable, the Lagrangian can be written as

\begin{eqnarray}\label{opt1}
&& \mathscr{L}(u,v,w,p,v_w,u^a,v^a,w^a,p^a,s)   \nonumber  \\
&=& \frac{\sigma_2}{2} \int_{X_0}^{X_1} \int_{\Gamma} |\frac{\partial v_w}{\partial X}|^2 + |v_w|^2 d\Gamma dX  \nonumber \\
&-&  \int_{X_0}^{X_1} \int_{\Gamma} s \left( v - v_w \right) d\Gamma dX
- \int_{0}^{X_0} \int_{\Gamma} s v d\Gamma dX
- \int_{X_1}^{X_t} \int_{\Gamma} s v  d\Gamma dX + \mathscr{H}
\end{eqnarray}
where the norm $\| \|$ is replaced by the absolute value since $v_w$ and $\tau_w$ are real functions. If we take the directional derivative of the Lagrangian with respect to $v_w$, the result is

\begin{equation}\label{ki}
\int_{X_0}^{X_1} \int_{\Gamma} \left[ \frac{\partial v_w}{\partial X} (\delta v_w)_X +  v_w \delta v_w  + s \delta v_w \right] d\Gamma dX = 0
\end{equation}
Then, the application of an integration by parts in $[X_0,X_1]$ about the first term in equation (\ref{ki}) yields

\begin{eqnarray}\label{}
&& \int_{\Gamma} \left[ \frac{\partial v_w}{\partial X} \delta v_w |_{X_1} -  \frac{\partial v_w}{\partial X} \delta v_w |_{X_0}  \right] d\Gamma  \nonumber \\
&+& \int_{X_0}^{X_{1}} \int_{\Gamma} \left[ -\frac{\partial^2 v_w}{\partial X^2} \delta v_w+  v_w \delta v_w  + s \delta v_w \right] d\Gamma dX = 0
\end{eqnarray}
so the optimality equation for $v_w$ is obtained in the form

\begin{equation}\label{eee}
\frac{\partial^2 v_w}{\partial X^2} - v_w = \frac{1}{\sigma_2} \left( p^a + \frac{\partial v^a}{\partial Y}  \right),
\end{equation}
satisfying the boundary conditions

\begin{equation}\label{bbb}
\frac{\partial v_w}{\partial X}(X_0) = 0, \frac{\partial v_w}{\partial X}(X_1) = 0,
\end{equation}
where $s$ is given by equation (\ref{ssa}), with $u=\hat{v}=w=0$ at the wall, and $\mathscr{F}=0$.

\section{APPENDIX C: Optimality condition for control based on wall deformation}

In the case of control based on wall deformation, the Lagrangian is given as

\begin{eqnarray}\label{opt2}
 \mathscr{L}(u,v,w,p,\mathscr{F},u^a,v^a,w^a,p^a,s) 
= \frac{\sigma_1}{2} \int_{X_0}^{X_1} \int_{\Omega} |\frac{\partial \mathscr{F}}{\partial X}|^2 + |\mathscr{F}|^2 d\Omega dX 
 + \mathscr{H}
\end{eqnarray}
where there is no component associated with boundary conditions since the wall deformation enters the problem through a Prandtl transformation (the functional $\mathscr{F}$ describing the wall deformations is a multiplicative factor in the state equations).

By taking the directional derivative with respect to $\mathscr{F}$, the optimality condition in this case is obtained as

\begin{eqnarray}\label{mmm}
&& \sigma_1 \frac{\partial^2 \mathscr{F}}{\partial X^2} - 2 \left( \frac{\partial u^a}{\partial z} \frac{\partial^2 u}{\partial Y^2} + \frac{\partial v^a}{\partial z} \frac{\partial^2 v}{\partial Y^2} + \frac{\partial w^a}{\partial z} \frac{\partial^2 w}{\partial Y^2} \right) \frac{\partial \mathscr{F}}{\partial z}  - \sigma_1 \mathscr{F}  \nonumber \\
&=& \left( \frac{\partial^2 u^a}{\partial z^2} \frac{\partial u}{\partial Y} + \frac{\partial^2 v^a}{\partial z^2} \frac{\partial v}{\partial Y} + \frac{\partial^2 w^a}{\partial z^2} \frac{\partial w}{\partial Y} \right)
- 2 \left( \frac{\partial u^a}{\partial z} \frac{\partial^2 u}{\partial Yz} + \frac{\partial v^a}{\partial z} \frac{\partial^2 v}{\partial Yz} + \frac{\partial w^a}{\partial z} \frac{\partial^2 w}{\partial Yz} \right)  \nonumber \\
&-& \left( u \frac{\partial u^a}{\partial X} + w \frac{\partial u^a}{\partial z} \right) \frac{\partial u}{\partial Y} -  \left( u \frac{\partial v^a}{\partial X} + w \frac{\partial v^a}{\partial z} \right) \frac{\partial v}{\partial Y} -  \left( u \frac{\partial w^a}{\partial X} + w \frac{\partial w^a}{\partial z} \right) \frac{\partial w}{\partial Y} \nonumber \\
&-& \frac{\partial p^a}{\partial X} \frac{\partial u}{\partial Y} - \frac{\partial p^a}{\partial z} \frac{\partial w}{\partial Y}+ \frac{\partial w^a}{\partial z} \frac{\partial p}{\partial Y}
\end{eqnarray}
subjected to boundary conditions

\begin{equation}\label{}
\frac{\partial \mathscr{F}}{\partial X}(X_0) = 0, \frac{\partial \mathscr{F}}{\partial X}(X_1) = 0.
\end{equation}
Equation (\ref{mmm}) is applied at the wall, where $u=v=w=0$ and $u_a=v_a=w_a=0$, except $u_a = \alpha (u_Y - \tau_0)$ in the interval $[X_{s0},X_{s1}]$. Thus, the optimality equation becomes:

\begin{eqnarray}\label{}
 \sigma_1 \frac{\partial^2 \mathscr{F}}{\partial X^2} - 2 \frac{\partial u^a}{\partial z} \frac{\partial^2 u}{\partial Y^2}  \frac{\partial \mathscr{F}}{\partial z}  - \sigma_1 \mathscr{F}
= \frac{\partial^2 u^a}{\partial z^2} \frac{\partial u}{\partial Y}- 2 \frac{\partial u^a}{\partial z} \frac{\partial^2 u}{\partial Yz}  - \frac{\partial p^a}{\partial X} \frac{\partial u}{\partial Y} - \frac{\partial p^a}{\partial z} \frac{\partial w}{\partial Y}
\end{eqnarray}

\section{APPENDIX D: Sensitivity of the results to the stream-wise resolution}\label{appD}

\begin{figure}
 \begin{center}
   \includegraphics[width=6cm]{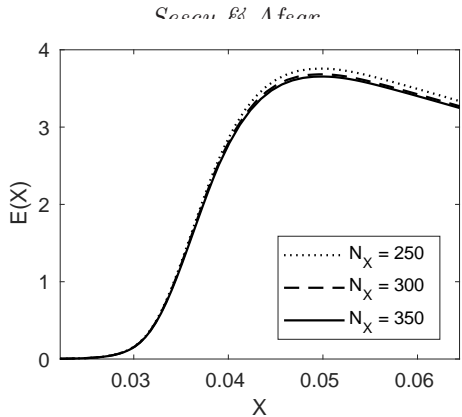} 
 \end{center}
  \caption{Vortex energy as a function of $X$ calculated using different grid spacings in the X direction.}
  \label{D}
\end{figure}

In figure \ref{D}, we show the sensitivity of vortex energy distribution as a function of $X$ to the grid spacings in the stream-wise direction ($N_x$ is the number of grid points in this direction).

\bibliography{jfm-references}
\bibliographystyle{jfm}

\end{document}